\documentclass[lettersize,journal]{IEEEtran}
\usepackage{amsmath,amsfonts}
\usepackage{algorithmic}
\usepackage{algorithm}
\usepackage{array}
\usepackage[caption=false,font=normalsize,labelfont=sf,textfont=sf]{subfig}
\usepackage{textcomp}
\usepackage{stfloats}
\usepackage{url}
\usepackage{verbatim}
\usepackage{graphicx}
\usepackage{cite}
\usepackage{xspace} 
\usepackage{amssymb}
\usepackage{multirow}
\usepackage{calc}
\usepackage{hyperref}
\newcommand{\eat}[1]{}
\newcommand{\ie}{\emph{i.e.,}\xspace}
\newcommand{\eg}{\emph{e.g.,}\xspace}
\newcommand{\aka}{\emph{a.k.a.}\xspace}
\newcommand{\etal}{\emph{et al.}\xspace}
\newcommand{\ours}{\textsc{LPO4Rec}\xspace}
\newcommand{\paratitle}[1]{\smallskip\noindent \textbf{#1}}

\begin{document}

\title{Listwise Preference Alignment Optimization for Tail Item Recommendation}

\author{Zihao Li, Chao Yang, Tong Zhang, Yakun Chen, Xianzhi Wang,~\IEEEmembership{Member,~IEEE}, Guandong Xu,~\IEEEmembership{Member,~IEEE}, Daoyi Dong,~\IEEEmembership{Fellow,~IEEE}
\thanks{Zihao Li, Tong Zhang, Yakun Chen, Xianzhi Wang and Daoyi Dong are with the Australian Artificial Intelligence Institute (AAII) and School of Compute Sicence, Faculty of Engineering and Information Technology, University of Technology Sydney, Sydney, NSW 2007, Australia (e-mails: zihao.li@student.uts.edu.au; tong.zhang-7@student.uts.edu.au; yakun.chen@student.uts.edu.au; xianzhi.wang@uts.edu.au; daoyi.dong@uts.edu.au)}
\thanks{Chao Yang is with the Faculty of Information Science and Engineering, Ocean University of China, Qingdao, China (e-mail:chao.yang@ouc.edu.cn)}
\thanks{Guandong Xu is with the Education University of Hong Kong, Hong Kong, China (e-mail: gdxu@eduhk.hk)}%
}

\markboth{Journal of \LaTeX\ Class Files,~Vol.~14, No.~8, August~2025}%
{Shell \MakeLowercase{\textit{et al.}}: A Sample Article Using IEEEtran.cls for IEEE Journals}


\maketitle

\begin{abstract}
Preference alignment has achieved greater success on Large Language Models (LLMs) and drawn broad interest in recommendation research.
Existing preference alignment methods for recommendation either require explicit reward modeling or only support pairwise preference comparison. The former directly increases substantial computational costs, while the latter hinders training efficiency on negative samples.
Moreover, no existing effort has explored preference alignment solutions for tail-item recommendation.
To bridge the above gaps, we propose \ours, which extends the Bradley-Terry model from pairwise comparison to listwise comparison, to improve the efficiency of model training.
Specifically, we derive a closed-form optimal policy to enable more efficient and effective training without explicit reward modeling. 
We also present an adaptive negative sampling and reweighting strategy to prioritize tail items during optimization and enhance performance in tail-item recommendations. 
Besides, we theoretically prove that optimizing the listwise preference optimization (LPO) loss is equivalent to maximizing the upper bound of the optimal reward.
Our experiments on three public datasets show that our method outperforms 10 baselines by a large margin, achieving up to $50\%$ performance improvement while reducing $17.9\%$ GPU memory usage when compared with direct preference optimization (DPO) in tail-item recommendation.
Our code is available at \url{https://github.com/Yuhanleeee/LPO4Rec}.
\end{abstract}


\section{Introduction}

\begin{figure}[!t]
\centering
\includegraphics[width=\columnwidth]{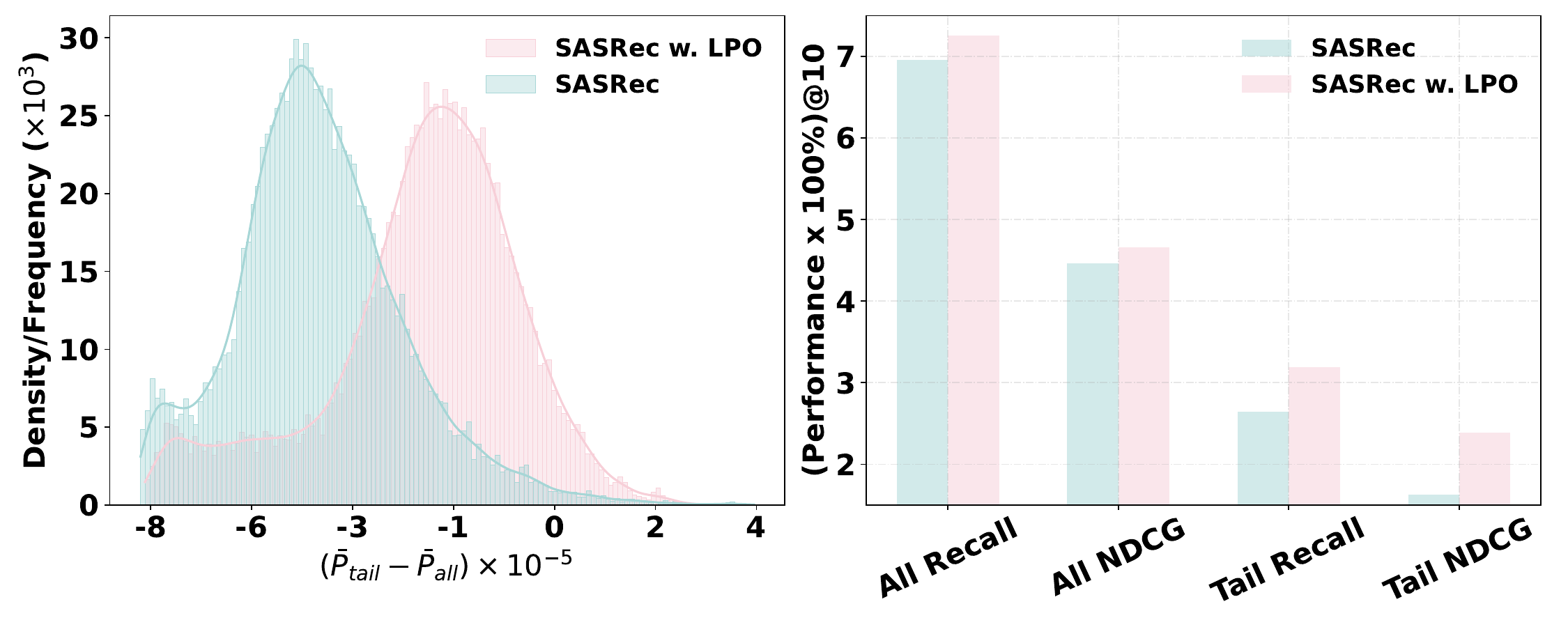}
\caption{Performance of SASRec and SASRec with our LPO loss function on the \textit{Amazon Beauty} dataset. The left subfigure illustrates the distributions of the differences in user-averaged recommendation probabilities of tail items and all items, \ie $\bar{P}_{tail}-\bar{P}_{all}$, showing that incorporating the LPO loss increases the probabilities of recommending tail items.
The right subfigure illustrates Recall@10 and NDCG@10 on all items and tail items, respectively, on the \textit{Amazon Beauty} dataset. Our approach exhibits superior performance over vanilla SASRec across all metrics.}
\label{fig:toy_exp}
\end{figure}

\IEEEPARstart{T}{he} success of Large Language Models (LLMs) in broad practical scenarios has inspired growing research on applying LLM-related techniques to recommendation tasks~\cite{yang2024harnessing,zhao2024recommender,bao2023tallrec}.
Preference alignment~\cite{bai2024aligning,li2024align,deng2025onerec} aims to align LLMs' behaviors with human preferences, which is crucial for enhancing LLMs' utility and compliance with the {3H principles} for responsible AI systems: \textit{Helpfulness}, \textit{Harmlessness}, and \textit{Honesty}~\cite{jiang2024survey,ouyang2022training}.
Existing preference alignment methods focused on bridging the inconsistency between LLMs and recommendation tasks in optimization objectives, and they have been actively studied for recommendation tasks~\cite{harte2023leveraging, zheng2024harnessing}.
Current preference alignment research employs either explicit or implicit reward modeling. Explicit reward modeling methods, \eg PPO~\cite{schulman2017proximal}, GRPO~\cite{shao2024deepseekmath}, and DAPO~\cite{yu2025dapo}, require training an explicit reward model to obtain a response preference score. 
Despite their advantages in handling complicated and logical reasoning tasks, these methods require learning at least two models simultaneously: a policy model and a reference model, causing significant efficiency issues.
Implicit reward modeling methods, \eg DPO~\cite{rafailov2023direct}, SimPO~\cite{meng2024simpo}, and ORPO~\cite{hong2024reference}, bypass intermediate steps to enable a more streamlined and efficient training process. Specifically, they eliminate the reward model by optimizing the policy model under the Bradley-Terry preference comparison framework~\cite{bradley1952rank}, enabling more stable and efficient fine-tuning of LLMs.
Despite elevated efficiency, they require pairwise comparisons of user preferences (\ie accepted sample vs. rejected sample), which limit their efficiency in learning from multiple rejected samples.
In addition, all existing preference alignment solutions for recommendation tasks rely on LLM frameworks~\cite{chen2024softmax, bai2024aligning}, which require extra training time and memory footprint, hindering their adoption in practical scenarios.
%
Furthermore, few studies have explored the potential of preference alignment methods for tail-item recommendation.

To bridge the aforementioned research gaps, we exploit the merits of preference alignment to improve the performance and diversity of tail-item recommendations.
Specifically, we incorporate the preference alignment optimization objective function into a sequential recommendation framework, targeting two objectives: (i) \textit{achieving efficient yet effective preference comparison optimization} and (ii) \textit{adapting the preference alignment strategy to enhance the performance of tail-item recommendation}.
To this end, we first extend the Bradley-Terry model in DPO from pairwise to listwise preference comparison.
Then, we derive the closed form of the reward model under its optimal policy, thereby proposing \ours (\ie \textbf{\underline{L}}istwise \textbf{\underline{P}}reference \textbf{\underline{O}}ptimization or LPO) and integrating it into the recommendation framework for model training.
LPO enables the comparison and optimization of multiple negative samples each time, allowing a more exhaustive exploitation of negative samples for an efficient preference alignment.
Finally, recognizing the significance of negative item sampling for output diversity in preference alignment~\cite{shi2023theories, ma2024negative}, we propose 1) an adaptive negative sampling strategy to align model output preferences with tail items; and 2) an adaptive reweighting strategy to dynamically adjust the sampling distribution of negative items based on the model optimization states.
Both strategies help direct the model’s focus toward prioritizing tail item optimization, thus improving recommendation performance on tail items.
Following SimPO and ORPO, we remove the reference model to further reduce the computational cost.

Our theoretical analysis shows that optimizing the LPO is equivalent to optimizing the upper bound of the optimal policy model under the extended listwise Bradley-Terry preference comparison framework.
LPO objection function also stands out in outputting diverse recommendations, facilitating tail-item recommendation.
Fig.~\ref{fig:toy_exp} shows our investigation of the recommendation probability distributions\footnote{Following previous research, we sort the items by the number of interactions and identify tail items based on the Pareto principle~\cite{box1986analysis}, \ie the $20/80$ rules.} and the performance of vanilla SASRec and the SASRec incorporated with our LPO loss on \textit{Amazon Beauty} dataset~\cite{ni2019justifying}.
Specifically, Fig.~\ref{fig:toy_exp} (left) investigates the difference in average recommendation probabilities between tail items and all items across all users.
It shows that incorporating our LPO loss increases the probability of recommending tail items, as indicated by the peak of the distribution of vanilla SASRec (colored in green) shifting closer to $0$ after being incorporated with our LPO loss (colored in pink).
Fig.~\ref{fig:toy_exp} (right) further shows that the LPO loss consistently enhances the performance of SASRec in both overall and tail-item recommendation scenarios.
In a nutshell, this paper makes the following main contributions:
\begin{itemize}
    \item We first introduce preference alignment optimization methods from LLM-based research to a general recommendation framework to deliver an efficient sequential recommendation approach for tail-item recommendation.
    \item We extend the Bradley-Terry preference comparison model from a pairwise to a listwise paradigm, to accelerate exploitation of negative samples and model training. We theoretically prove that optimizing the LPO is equivalent to optimizing the upper bound of the optimal policy model, and our LPO promotes output diversity to enhance tail-item recommendation. 
    \item We propose an adaptive negative sampling strategy, which dynamically allocates preference priorities to tail items based on their optimization, navigating the model's focus toward these items. 
    \item Our extensive experiments demonstrate that our method achieves up to a $50\%$ performance improvement in tail-item recommendation, with only $82.1\%$ memory usage, compared to the representative preference alignment optimization algorithm, DPO.
\end{itemize}

This paper is organized as follows. Section~\ref{sec:related} presents a brief introduction to related work. Section~\ref{sec:preliminary} outlines the sequential recommendation task and the DPO method for preference alignment. Building on these preliminaries, Section~\ref{sec:meth} extends DPO for list-wise preference alignment, further, adaptive negative sampling and adaptive loss reweighting are proposed to enhance tail-item recommendation. In Section~\ref{sec:exp}, we compare our method with ten baselines in terms of both performance and training efficiency. Finally, Section~\ref{sec:conclusion} concludes the paper.

\section{Related Work}
\label{sec:related}
\subsection{Sequential Recommendation}

\paratitle{Conventional Methods.} Sequential recommendation aims to predict the next preferred items based on users' chronologically ordered history records. 
Sequential recommendation has become ubiquitous across various domains, thanks to its superior capability to capture interaction patterns among items, such as e-commerce, social media, and streaming services.
Conventional sequential recommendation methods apply the Markov chain, which recognizes users' sequential interaction behaviors as a Markov Decision Process (MDP) for next-item prediction~\cite{shani2005mdp, rendle2010factorizing}.
Deep neural networks have been widely applied to recommendation tasks~\cite {zhang2019deep, li2024graph}, given their success in various practical applications~\cite{pouyanfar2018survey, lecun2015deep, yang2025deep}.
In particular, solutions based on Convolutional Neural Network (CNN)~\cite{tang2018personalized}, Gated Recurrent Unit (GRU)~\cite{hidasi2015session}, Variational Autoencoder (VAE)~\cite{xie2021adversarial}, Graph Neural Network, Transformer~\cite{kang2018self,sun2019bert4rec} and its variants~\cite{fan2021continuous,wu2020sse,xia2022multi} have been proposed for sequential recommendation.
Reinforcement learning~\cite{wang2020kerl} and diffusion methods~\cite{li2023diffurec, wang2023diffusion} are also explored for sequential recommendation. 

\paratitle{LLM-based Methods.}
LLMs have been increasingly applied to sequential recommendation~\cite{harte2023leveraging, zheng2024harnessing}.
LLM-based methods generally fall into two categories: \textit{LLM-as-encoder} and \textit{LLM-as-recommender}. 
LLM-as-encoder solutions~\cite {hou2022towards, li2023text} differ from conventional approaches in leveraging large language models to encode the associated text information (\eg title, description) of items as an alternative to ID-based embedding for representation generation.
In contrast, LLM-as-recommender solutions identify the recommendation task (\ie next-item prediction) straightforwardly as a next-token prediction and leverage a unified text-to-text generation paradigm with prompt fine-tuning for sequential recommendation~\cite{geng2022recommendation, shu2024rah}.
Other techniques following this idea include parameter-efficient fine-tuning~\cite{peng2023towards}, retrieval argumentation~\cite{zhao2024raserec}, and multi-modality modeling~\cite{ye2025harnessing}.
Compared with previous methods, our \ours features in being model-agnostic, efficient, and suitable to be integrated into model training (based on a dedicated loss function) for performance improvement.

\subsection{Tail-item Recommendation Enhancement}

In practice, users' attention and interactions are primarily focused on popular (head) items, causing head items to dominate recommendation outputs.
In contrast, the vast majority of items, \aka tail items, receive minimal exposure, compromising diversity in outcomes and fairness of recommendation systems~\cite{yin2012challenging, li2017two}. 
Representative strategies to address the tail-item challenge include \textit{resampling, reweighting}~\cite{huang2006correcting, li2025reembedding} and \textit{knowledge transfer}~\cite{zhang2021model}.
Reweighting and resampling approaches amplify the optimization focus on tail items with tailored weight functions during model training, steering the model to attend more to tail items to enhance their chance to be recommended. 
Knowledge transfer methods augment tail-item representations by leveraging auxiliary information, such as collaborative signals from head items~\cite{liu2023co, kim2023melt, jang2020cities, qin2024metaga}, similar items~\cite{yang2023loam, wang2023multifdf}, sequential patterns~\cite{hu2022memory}, high-level semantic representations derived from LLMs~\cite{liu2024large}, LVMs~\cite{bian2023multi}, and knowledge graphs~\cite{zhang2024relation}.
Our \ours not only leverages resampling and reweighting strategies for tail-item enhancement but also explores those strategies from the novel and emerging perspective of preference alignment, which was not considered in existing research.

\subsection{Preference Alignment with LLMs for Recommendation}

Preference alignment optimization aims to align LLMs' outputs to human preferences to ensure \textit{harmlessness}, \textit{helpfulness}, and \textit{honesty} ~\cite{bai2022constitutional} of recommendation results.
Many studies apply preference alignment optimization for controllable generation~\cite{guo2024controllable, wang2024arithmetic}. 
Reinforcement Learning from Human Feedback (RLHF)~\cite{ouyang2022training} is a pioneer work, which explicitly defines the reward model with the PPO training algorithm for preference alignment. 
On this basis, DPO~\cite{rafailov2023direct} derives the closed form of optimal reward from the optimal policy and is introduced into the Bradley-Terry comparison model for preference alignment.
SimPO~\cite{meng2024simpo} and ORPO~\cite{hong2024reference} eliminate the reference model during training to overcome the bottleneck of the exorbitant computation cost of model training.
Regarding LLM-based recommendation research, tailored preference alignment optimization is a cutting-edge area.
For example, Bai \etal~\cite{li2024align} extended the DPO algorithm with negative sampling to support multiple negative samples preference alignment and recommendation. Deng \etal~\cite{deng2025onerec} proposed an iterative preference alignment algorithm under the DPO, which generates the negative samples based on a learnable reward model.
In this way, the model can be trained iteratively based on the DPO algorithm.
Chen \etal~\cite{chen2024softmax} devised a softmax-DPO (S-DPO) as an alternative to the standard DPO algorithm, which incorporates multiple negatives to support partial ranking for LLM-based recommendation.
Our work distinguishes itself from the above studies in proposing a listwise preference alignment algorithm with tailored adaptive negative sampling and reweighting strategies for efficient and effective tail-item recommendation.
We also provide the theoretical proof of the advantages of our approach in mining hard samples and enhancing recommendation diversity.

\section{Preliminaries}
\label{sec:preliminary}

\subsection{Sequential Recommendation}

Sequential recommendation aims to capture the dynamic evolution and temporal behavioral patterns of user preferences, including short-term preference and long-term tendencies, for next-item prediction. 
Given a user $u$, whose interaction history is sorted chronologically into a sequence \ie $\mathcal{S}_u=[x_{1}, x_{2}, ..., x_{t}]$, where $t$ is a timestamp, sequential recommendation aims to predict the probability $P$ of items that the user is likely to engage with at the subsequent time step $t+1$.
We formalize the prediction task as:
\begin{equation}
P(x_{t+1}|x_1,x_2,...x_t)=f_\theta(x_1,x_2,...,x_t)\label{eq:sq},
\end{equation}
where $f_{\theta}(\cdot)$ denotes the parameterized recommendation model responsible for next-item prediction.

\subsection{Direct Preference Optimization}

\paratitle{Reinforcement Learning from Human Feedback (RLHF).}
Built upon the foundational work~\cite{jaques2017sequence}, preference alignment with reinforcement learning requires explicitly defining the reward model for human feedback assessment. The optimization objective is formalized as:
\begin{equation}
    \max_{\pi_\theta} \mathbb E_{x\sim D, y\sim \pi_{\theta}(y|x)}[r(x,y)]-\beta \mathbb{D}_{\textrm{KL}}[\pi_\theta(y|x)||\pi_{\textrm{ref}}(y|x)],
\label{eq:rlhf}
\end{equation}
where $r(\cdot)$ denotes the reward model. $\beta$ controls the deviation of the language model policy $\pi_{\theta}(\cdot)$ from the base reference policy $\pi_{\textrm{ref}}(\cdot)$. $\mathbb{D}_{\textrm{KL}}(P(a)||P(b))$ calculates the KL-divergence between the distribution of $P(a)$ and $P(b)$.

\paratitle{The Closed Form of Optimal Reward Function.} Following~\cite{rafailov2024direct}, we derive the closed form of the reward model $r(x,y)$ from Eq~\eqref{eq:rlhf}:
\begin{equation}
\begin{aligned}
    &\max_{\pi_\theta} \mathbb E_{x\sim D, y\sim \pi_{\theta}(y|x)}[r(x,y)]-\beta \mathbb{D}_{\textrm{KL}}[\pi_\theta(y|x)||\pi_{\textrm{ref}}(y|x)]\\
    &=\max_{\pi_\theta} \mathbb E_{x\sim D}\mathbb{E}_{y\sim \pi_{\theta}(y|x)}\left[r(x,y)-\beta \textrm{log}\frac{\pi_{\theta}(y|x)}{\pi_{\textrm{ref}}(y|x)}\right]\\
    &=\min_{\pi_\theta} \mathbb E_{x\sim D}\mathbb{E}_{y\sim \pi_{\theta}(y|x)}\left[\textrm{log}\frac{\pi_{\theta}(y|x)}{\pi_{\textrm{ref}}(y|x)}-\frac{1}{\beta}r(x,y) \right]\\
    &= \min_{\pi_\theta} \mathbb E_{x\sim D}\, \mathbb{E}_{y\sim \pi_{\theta}(y|x)}
    \mathopen{\Bigg[}
      \log\frac{\pi_{\theta}(y|x)}{\dfrac{1}{Z(x)}\pi_{\textrm{ref}}(y|x)\exp\left(\frac{1}{\beta}r(x,y)\right)} 
      \vphantom{\dfrac{1}{Z(x)}} \\
    &\quad - \log Z(x)
    \mathclose{\Bigg]} ,
\end{aligned}
\label{eq:rw}
\end{equation}
where $Z(x)=\sum_y\pi_{\textrm{ref}}(y|x)\textrm{exp}\left(\frac{1}{\beta}r(x,y)\right)$ is a partition function. 

Consistent with the results established in~\cite{peters2010relative, peng2019advantage}, we can obtain an exact analytical solution:
\begin{equation}
    \pi_{\theta}(y|x)=\frac{1}{Z(x)}\pi_{\textrm{ref}}(y|x)\textrm{exp}\left(\frac{1}{\beta}r(x,y)\right).
    \label{eq:pi}
\end{equation}

While the solution to Eq.~\eqref{eq:pi} in itself is intractable, the optimal reward can be expressed as a function in terms of its corresponding optimal policy $\pi_{\theta}^*(y|x)$, as formalized below: 
\begin{equation}
    r^*(x,y) = \beta \textrm{log}\frac{\pi^*_\theta(y|x)}{\pi_{\textrm{ref}}(y|x)}+ \beta \textrm{log} Z(x).
    \label{eq:reward_opt}
\end{equation}

\paratitle{Bradley-Terry preference model.} The Bradley-Terry model is a probabilistic framework for analyzing the outcome of pairwise comparisons between objects. Given two outputs within the context of the reward model for preference comparison, $y_1$ and $y_2$, the Bradley-Terry model can be reformulated as:
\begin{equation}
\begin{aligned}
    p^*(y_1\succ y_2|x)&=\frac{\textrm{exp}(r^*(x,y_1))}{\textrm{exp}(r^*(x,y_1))+\textrm{exp}(r^*(x,y_2))}\\
    &=\sigma(r^*(x,y_1)-r^*(x,y_2)),
\end{aligned}
    \label{eq:bt}
\end{equation}
where $\sigma(\cdot)$ is the sigmoid function.

\paratitle{Deriving the DPO.} Substituting the optimal reward function Eq.~\eqref{eq:reward_opt} into the Bradley-Terry preference model Eq.~\eqref{eq:bt}, the partition function $Z(x)$ can be canceled out. Therefore, we have:
\begin{equation}
\begin{aligned}
    &p^*(y_1\succ y_2|x)=\\&\frac{\textrm{exp}\left(\beta \textrm{log}\frac{\pi^*_\theta(y_1|x)}{\pi_{\textrm{ref}}(y_1|x)}+ \beta \textrm{log} Z(x) \right)}{\textrm{exp}\!\left( \beta \textrm{log}\frac{\pi^*_\theta(y_1|x)}{\pi_{\textrm{ref}}(y_1|x)} \!\!+\!\!\beta \textrm{log} Z(x)\right)\!\!+\!\! \textrm{exp}\!\left(\beta \textrm{log}\frac{\pi^*_\theta(y_2|x)}{\pi_{\textrm{ref}}(y_2|x)}\!\!+\!\!\beta \textrm{log} Z(x) \right)}\\
    &= \frac{1}{1+\textrm{exp}\left(\beta\textrm{log}\frac{\pi^*(y_2|x)}{\pi_{\textrm{ref}}(y_2|x)}- \beta\textrm{log}\frac{\pi^*(y_1|x)}{\pi_{\textrm{ref}}(y_1|x)}\right) }\\
    &=\sigma\left( \beta \textrm{log}\frac{\pi^*(y_1|x)}{\pi_{\textrm{ref}}(y_1|x)}- \beta \textrm{log} \frac{\pi^*(y_2|x)}{\pi_{\textrm{ref}}(y_2|x)} \right).
\end{aligned}
\end{equation}

Consequently, the DPO loss objection can be formalized as:
\begin{equation}
\begin{aligned}
    &\mathcal{L}_\textrm{DPO}(\pi_\theta;\pi_{\textrm{ref}})\\
    &=-\mathbb{E}_{(x,y_1,y_2)\sim\mathcal{D}}\left[\textrm{log}\sigma\left(\beta \textrm{log}\frac{\pi_\theta(y_1|x)}{\pi_{\textrm{ref}}(y_1|x)}- \beta \textrm{log}\frac{\pi_\theta(y_2|x)}{\pi_{\textrm{ref}}(y_2|x)} \right) \right],
    \label{eq:dpo}
\end{aligned}
\end{equation}
where $y_1$ and $y_2$ denote the accepted and rejected output pairs, given the input $x$. Optimizing the policy $\pi_{\textrm{ref}}$ is equal to optimizing the optimal reward under the $\mathcal{L}_{DPO}$ loss.

\section{Methodology}
\label{sec:meth}

\subsection{Listwise Preference Optimization (LPO) Loss}
\label{sec:lpo}

\begin{figure*}[!t]
\centering
\includegraphics[width=\textwidth]{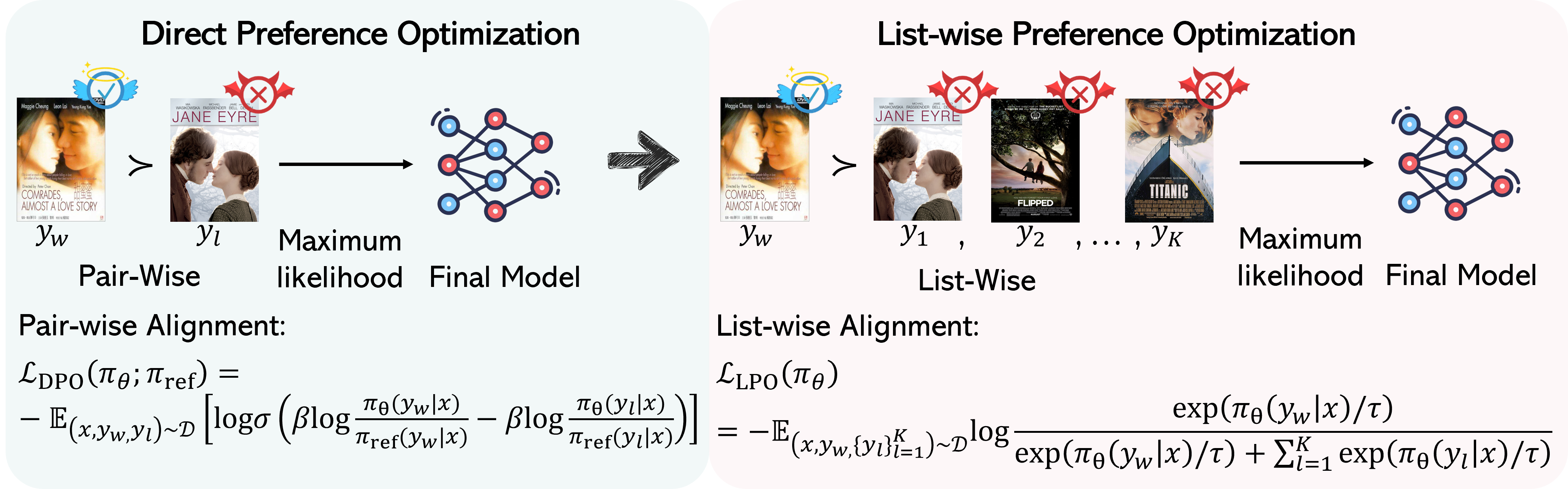}
\caption{Comparison of DPO (left) and LPO (right) loss functions. DPO optimizes the model with pairwise ($y_w$ vs. $y_\ell$) comparison for preference alignment, where both the policy $\pi_{\theta}(\cdot)$ and reference $\pi_{\textrm{ref}}(\cdot)$ models are involved in the training process.
In contrast, our LPO loss leverages listwise ($y_w$ vs. $[y_{\ell_1}, y_{\ell_2}, ..., y_{\ell_K}]$) comparison and remove the reference model, where $y_w$ and $y_\ell$ are accepted (positive) and rejected (negative) samples, respectively}
\label{fig:framework}
\end{figure*}

Constrained by the Bradley-Terry model, the vanilla DPO loss (Eq.~\eqref{eq:dpo}) only allows pairwise sample comparison, causing inefficiency in handling multiple negative samples.
We derive a generalized form of the Bradley-Terry model by extending the Bradley-Terry preference model from pairwise to listwise comparison.
Given input $x$ and multiple outputs $[y_1,y_2,...,y_K]$, we reformulate the Bradley-Terry model by considering the mutual independence of the probabilities associated with any two outputs $y_i$ and $y_j$:
\begin{equation}
\begin{aligned}
    &p\left(y_w\succ \{y_\ell\}_{\ell=1}^K|x\right)\\&=\prod_{\ell=1}^K p(y_w\succ y_\ell|x)
    =\prod_{\ell=1}^K\frac{\textrm{exp} (r(y_w,x))}{\textrm{exp} (r(y_w,x))+\textrm{exp}(r(y_\ell,x))},
    \label{eq:bt_list}
\end{aligned}
\end{equation}
where $y_w$ and $y_{\ell}\in\{y_1,y_2,...,y_K\}$ denote the accepted and rejected samples, respectively; $K$ is the number of rejected samples; $r(y,x)$ denotes the reward score.

Similar to DPO, by substituting Eq.~\eqref{eq:reward_opt} into Eq.~\eqref{eq:bt_list}, we have:
\begin{equation}
    \begin{aligned}
       &p^*\left(y_w\succ \{y_\ell\}_{\ell=1}^K|x\right)\\
       &=\prod_{\ell=1}^K\frac{\textrm{exp}\left(\beta \textrm{log}\frac{\pi^*_\theta(y_w|x)}{\pi_{\textrm{ref}}(y_w|x)}+ \beta \textrm{log}Z(x)\right)}{\textrm{exp}\left(\beta \textrm{log}\frac{\pi^*_\theta(y_w| x)}{\pi_{\textrm{ref}}(y_w| x)}\!\!+\!\!\beta \textrm{log}Z(x)\right) \!\!+\!\! \textrm{exp}\left(\beta \textrm{log}\frac{\pi^*_\theta(y_\ell| x)}{\pi_{\textrm{ref}}(y_\ell| x)}\!\!+\!\!\beta \textrm{log}Z(x) \right)}\\
       &=\prod_{\ell=1}^K\frac{\textrm{exp}\left(\beta \textrm{log}\frac{\pi^*_\theta(y_w|x)}{\pi_{\textrm{ref}}(y_w|x)}\right)}{\textrm{exp}\left(\beta \textrm{log}\frac{\pi^*_\theta(y_w| x)}{\pi_{\textrm{ref}}(y_w| x)}\right) + \textrm{exp}\left(\beta \textrm{log}\frac{\pi^*_\theta(y_\ell| x)}{\pi_{\textrm{ref}}(y_\ell| x)} \right)}.
    \end{aligned}
    \label{eq:btl}
\end{equation}

The degree of preference between accepted and rejected samples, $\beta$, typically ranges from $0.05$ to $5$.
Here, we set $\beta=1$ for clarity of derivation and formalization.
Further, considering $\prod_{i=1}^K(1+x_i)>1+\sum_{i=1}^Kx_i$ when $x_i>0$, we re-formalize Eq.~\eqref{eq:btl} into:
\begin{equation}
\begin{aligned}
     &p^*\left(y_w\succ \{y_\ell\}_{\ell=1}^K|x\right)\
     =\prod_{\ell=1}^K\frac{\frac{\pi^*_\theta(y_w|x)}{\pi_{\textrm{ref}}(y_w|x)}}{\frac{\pi^*_\theta(y_w|x)}{\pi_{\textrm{ref}}(y_w|x)}+\frac{\pi^*_\theta(y_\ell|x)}{\pi_{\textrm{ref}}(y_\ell|x)}}\\
     &\propto\prod_{\ell=1}^K\frac{\textrm{exp}\left(\frac{\pi^*_\theta(y_w|x)}{\pi_{\textrm{ref}}(y_w|x)}\right)}{\textrm{exp}\left(\frac{\pi^*_\theta(y_w|x)}{\pi_{\textrm{ref}}(y_w|x)}\right)+\textrm{exp}\left(\frac{\pi^*_\theta(y_\ell|x)}{\pi_{\textrm{ref}}(y_\ell|x)}\right)}\\
     &=\prod_{\ell=1}^K\frac{1}{1+\textrm{exp}\left(\frac{\pi^*_\theta(y_\ell|x)}{\pi_{\textrm{ref}}(y_\ell|x)}-\frac{\pi^*_\theta(y_w|x)}{\pi_{\textrm{ref}}(y_w|x)} \right)}\\
     &<\frac{1}{1+\sum_{\ell=1}^K\textrm{exp}\left(\frac{\pi^*_\theta(y_\ell|x)}{\pi_{\textrm{ref}}(y_\ell|x)}-\frac{\pi^*_\theta(y_w|x)}{\pi_{\textrm{ref}}(y_w|x)} \right)}.
\end{aligned}
\label{eq:bound_listbt}
\end{equation}

Inspired by SimPO and ORPO, we further remove the reference model $\pi_{\textrm{ref}}(\cdot)$ to improve the training efficiency.
Finally, the LPO loss is derived from the maximum likelihood objective of Eq.~\eqref{eq:bound_listbt}, as formalized below:
\begin{equation}
\begin{aligned}
    &\mathcal{L}_{\textrm{LPO}}(\pi_\theta)\\
    &=-\mathbb{E}_{\left(x,y_w,\{y_\ell\}_{\ell=1}^K\right)\sim \mathcal{D}}\textrm{log}\frac{1}{1+\displaystyle \sum_{\ell=1}^K  \textrm{exp}\left(\pi_{\theta}(y_\ell|x)-\pi_{\theta}(y_w|x)\right)}\\
    &=\!\!-\!\mathbb{E}_{\left(x,y_w,\{y_\ell\}_{\ell=1}^K\right)\sim \mathcal{D}}\textrm{log}\frac{\textrm{exp}\left(\pi_\theta(y_w|x)/\tau\right)}{\textrm{exp}\!\left(\pi_\theta(y_w|x)/\tau\!\right)\!\!+\!\!\!\displaystyle \sum_{\ell=1}^K\textrm{exp}\!\left(\pi_\theta(y_\ell|x)/\tau\!\right)},
\end{aligned}
\label{eq:lpo}
\end{equation}
where $\tau$ is the temperature ratio, which controls the scaling of preference differences by adjusting the shape of the probability distribution. A smaller $\tau$ enlarges the preference gap between the accepted and rejected samples, yielding a sharper probability distribution. 
Comparing Eq.~\eqref{eq:lpo} with Eq.~\eqref{eq:bound_listbt}, we optimize $\mathcal{L}_{LPO}$ to closely approximate the upper bounds of the optimal policy model $\pi_{\textrm{ref}}^*$ within the Bradley-Terry framework.
A comparison between DPO with our LPO is shown in Fig.~\ref{fig:framework}.

\subsection{LPO for Sequential Recommendation}

Following a previous study~\cite{hong2024reference}, we jointly model sequential recommendation and preference alignment by integrating the cross-entropy (CE) loss with our LPO loss (Eq.~\eqref{eq:lpo}). 
Given a sequence $S_i=[x_1,x_2,...,x_t,x_{t+1}]$, suppose $x_{t+1}$ is the next item to recommend (\ie the ground-truth, denoted by $y_{w_i}$).
The completed loss function of $\pi_{\theta}$ is formalized as:
\begin{equation}
    \begin{aligned}
\mathcal{L}&=\mathcal{L}_{\textrm{CE}}+\lambda\mathcal{L}_{\textrm{LPO}}\\
&=-\frac{1}{N}\sum_{i=1}^N\textrm{log}\pi_{\theta}(y_{w_i}|S_i)+\\
&\lambda\textrm{log}\frac{\textrm{exp}\left(\pi_\theta(y_{w_i}|S_i)/\tau\right)}{\textrm{exp}\left(\pi_\theta(y_{w_i}|S_i)/\tau\right)+\sum_{\ell=1}^K\textrm{exp}\left(\pi_\theta(y_{\ell_i}|S_i)/\tau\right)},
    \end{aligned}
    \label{eq:lposft}
\end{equation}
where $N$ denotes the number of training samples. $y_{\ell_i}$ is a negative item~\footnote{In recommendation context, we refer to \textit{rejected samples} as \textit{negative samples}, and \textit{accepted samples} as \textit{positive samples} or \textit{ground-truth}.} corresponding to $y_{w_i}$. $\lambda$ is a hyperparameter controlling the weight of preference alignment loss relative to sequential recommendation loss.

\subsection{Tail-item Recommendation Augmentation}
\label{sec:sample}

\paratitle{Adaptive Negative Sampling.}
The optimization objective Eq.~\eqref{eq:lpo} requires a list of negative samples (\ie $y_{\ell}, \ell=1,2,...,K$) for preference alignment.
We design an adaptive negative sampling strategy to enhance the model's capability in tail-item recommendation by 1) prioritizing drawing negative samples from the head item set $\mathcal{I}_{H}$ and 2) updating the sampling probabilities of head items during model training, so that head items with higher predicted probabilities are more likely to be drawn as negative samples.
As such, the performance on tail items is improved by suppressing the recommendation probabilities of head items. 
The above negative sampling function is formalized as follows:
\begin{equation}
    y_{\ell}\sim P(y_{\ell})=\frac{\textrm{exp}(\pi_{\theta}(y_{\ell}|S))}{\sum_{y_{\ell}\in \mathcal{I}_{H}}\textrm{exp}(\pi_{\theta}(y_{\ell}|S))}, \; \ell=1,2,...,K,
    \label{eq:neg_sample}
\end{equation}

The Top-K sampling process is non-differentiable due to the presence of the $\mathsf{arg max}$ operation.
We adopt the Gumbel-Softmax~\cite{maddison2016concrete} trick to make it differentiable.
Specifically, a Gumbel noise $G_{y_{\ell}}$ drawn from the distribution $\textrm{Gumbel}(0, \beta)\sim-\beta\textrm{log}(-\textrm{log}(0,\beta))$ is added into the probability $\pi_{\theta}(y_{\ell}|S)$ as a perturbation for Top-K sampling~\cite{kool2019stochastic}, where $\beta=1, U\sim\textrm{Uniform}(0, 1)$.
In the forward path, the $\mathsf{arg max}$ operation is applied over the Gumbel softmax to sample Top-K negatives; in the backward path, gradients are computed through the continuous Gumbel-Softmax distribution.
The above process is formalized as:
\begin{equation}
    y_{\ell}\sim P(y_{\ell})=\frac{\textrm{exp}(\pi_{\theta}(y_{\ell}|S)+G_{y_{\ell}})}{\sum_{y_{\ell}\in \mathcal{I}_{H}}\textrm{exp}(\pi_{\theta}(y_{\ell}|S)+G_{y_{\ell}})}.
    \label{eq:neg_sample_gumbel}
\end{equation}

\paratitle{Adaptive Loss Reweighting.}
Inspired by~\cite{li2025reembedding}, we apply a reweighting strategy for further improve tail-item recommendation. 
The reweighting function aims to assign higher weights to samples whose predicted labels ($x_{t+1}$) belong to tail items, thus enhancing the optimization intensity on tail items during model training.
The reweighting function is defined as follows:
\begin{equation}
\begin{aligned}
    & \omega_i=\frac{\textrm{exp}(\alpha_i)}{\sum_{i=1}^m\textrm{exp}(\alpha_i)},
    &\alpha_i=\begin{cases}
        \alpha_H,\quad \textrm{if $x_{t+1}\in \mathcal{I_H}$}\\
        \alpha_T,\quad \textrm{if $x_{t+1}\in \mathcal{I_T}$},
    \end{cases}
\end{aligned} 
    \label{eq:reweight}
\end{equation}
where $m$ denotes the batch size. $\mathcal{I}_{T}$ and $\mathcal{I}_{H}$ denote the sets of tail and head items, respectively. 
Here, we define $\alpha_T=1, \alpha_H=0$, ensuring $\alpha_T>\alpha_T$---this allows the model to assign greater weight to samples from the tail item set during optimization.
Accordingly, the loss function Eq.~\eqref{eq:lposft} is updated as:
\begin{equation}
\mathcal{L}=\omega(\mathcal{L}_{\textrm{CE}}+\lambda \mathcal{L}_{\textrm{LPO}}).
\label{eq:loss_final}
\end{equation}

Algorithm~\ref{alg:lpo} describes the completed training process.

\subsection{Properties of LPO}
\label{sec:prop}

We analyze $\mathcal{L}_\textrm{LPO}$ by first deriving its gradient with respect to $\pi_{\theta}(y_{w}|x)$:
\begin{equation}
\begin{aligned}
    &\frac{\partial \mathcal{L}_\textrm{LPO}}{\partial \pi_{\theta}(y_w|x)}
    \!=\!\frac{\partial}{\partial \pi_{\theta}(y_{w}|x)}\!-\!\textrm{log}\frac{\textrm{exp}(\pi_{\theta}(y_{w}|x))}{\textrm{exp}(\pi_{\theta}(y_{w}|x))\!+\!\sum_{\ell=1}^K\!\textrm{exp}(\pi_{\theta}(y_{\ell}|x))}\\
    &=\frac{\partial}{\partial \pi_{\theta}(y_{w}|x)}\textrm{log}\left(\frac{\textrm{exp}(\pi_{\theta}(y_{w}|x))+\sum_{\ell=1}^K\textrm{exp}(\pi_{\theta}(y_{\ell}|x))}{\textrm{exp}(\pi_{\theta}(y_{w}|x))}\right)\\
    &=\frac{\textrm{exp}(\pi_{\theta}(y_{w}|x))}{\textrm{exp}(\pi_{\theta}(y_{w}|x))+\sum_{\ell=1}^K\textrm{exp}(\pi_{\theta}(y_{\ell}|x))}\cdot\\
    &\frac{\partial}{\partial \pi_{\theta}(y_{w}|x)}\frac{\textrm{exp}(\pi_{\theta}(y_{w}|x))+\sum_{\ell=1}^K\textrm{exp}(\pi_{\theta}(y_{\ell}|x))}{\textrm{exp}(\pi_{\theta}(y_{w}|x))}\\
    &=\frac{\textrm{exp}(\pi_{\theta}(y_{w}|x))}{\textrm{exp}(\pi_{\theta}(y_{w}|x))+\sum_{\ell=1}^K\textrm{exp}(\pi_{\theta}(y_{\ell}|x))}\cdot\\
    &\frac{-\sum_{\ell=1}^K\textrm{exp}(\pi_{\theta}(y_{\ell}|x))}{\textrm{exp}(\pi_{\theta}(y_{w}|x))^2}\cdot \textrm{exp}(\pi_{\theta}(y_{w}|x))\\
    &=-\frac{\sum_{\ell=1}^K\textrm{exp}(\pi_{\theta}(y_{\ell}|x))}{\textrm{exp}(\pi_{\theta}(y_{w}|x))+\sum_{\ell=1}^K\textrm{exp}(\pi_{\theta}(y_{\ell}|x))}\\
    &=\frac{\textrm{exp}(\pi_{\theta}(y_{w}|x))}{\textrm{exp}(\pi_{\theta}(y_{w}|x))+\sum_{\ell=1}^K\textrm{exp}(\pi_{\theta}(y_{\ell}|x))}-1.
\end{aligned}
\label{eq:grad_lpo}
\end{equation}

Based on the law of large numbers, we replace the sum operation over $y_{\ell}$ with an expectation. 
Note that $\textrm{exp}(\pi_{\theta}(y|x))\ll K\mathbb{E}_{y_{\ell}\sim p}\textrm{exp}(\pi_{\theta}(y_{\ell}|x))$, when $K$ is large. Therefore, the Eq.~\eqref{eq:grad_lpo} can be rewritten into
\begin{equation}
\begin{aligned}
     &\frac{\textrm{exp}(\pi_{\theta}(y_{w}|x))}{\textrm{exp}(\pi_{\theta}(y_{w}|x))+K\mathbb{E}_{y_{\ell}\sim p(y_{\ell}|x)}\textrm{exp}(\pi_{\theta}(y_{\ell}|x))}-1\\
     &\propto \frac{\textrm{exp}(\pi_{\theta}(y_{w}|x))}{K\mathbb{E}_{y_{\ell}\sim p(y_{\ell}|x)}\textrm{exp}(\pi_{\theta}(y_{\ell}|x))}-1.
\end{aligned} 
\label{eq:grad_property}
\end{equation}

In Eq.~\eqref{eq:grad_property}, the negative gradient increases with the decrease of the numerator $\pi_{\theta}(y_{w}|x)$ or the increase of denominator $\pi_\theta(y_{\ell}|x)$.
This brings about two merits of LPO: 1) \textit{focusing on hard-item optimization}. According to the numerator of Eq.~\eqref{eq:grad_property}, a smaller probability $\pi_{\theta}(y_w|x)$ amplifies the gradient magnitude. Therefore, the LOP loss encourages the model to prioritize the optimization of hard items (\ie ground truth with lower recommendation probabilities) during training. 2) \textit{promoting tail items and diversity of recommendation outputs}. According to the denominator of Eq.~\eqref{eq:grad_property}, negative samples $y_{\ell}$ are predominantly drawn from the head items, given that sampling probability is proportional to the recommendation probability, and head items generally have higher recommendation probabilities than tail items (Fig.~\ref{fig:case}).
Accordingly, the recommended positive items (\eg $y_{w}$) have a preference for tail items, promoting diversity and recommendations of tail items.

\begin{algorithm}[H]
\caption{Listwise Preference Alignment Optimization for Sequential Recommendation (\ours).}\label{alg:lpo}
\begin{algorithmic}
\STATE 
\STATE {\textsc{INPUT:}}
\STATE \textbf{Data:} User historical interacted sequence dataset $\mathcal{D}=\{(S,y_w)_i\}_{i=1}^N$ of size $N$, where $S_i=[x_1,x_2,...,x_t]$, $y_{w_i}=x_{t+1}$; Tail item set $\mathcal{I}_T$; Head item set $\mathcal{I}_H$
\STATE  \textbf{Hyperparameter:} Learning rate $\eta$; Training epoch $E$; Temperature ratio $\tau$; Rejected item numbers $K$; Coefficient $\lambda$; Reweighting factor $\alpha_T$ and $\alpha_H$
\STATE  \textbf{Model:} Policy model $\pi_{\theta}(\cdot)$
\STATE 
\STATE {\textsc{TRAINING:}}
\FOR{each epoch in $E$}
            \STATE \textbf{Batch sample construction:} Sample mini-batch $\mathcal{D}_m=\{(S,y_w)_i\}_{i=1}^m$ from $\mathcal{D}$\;
            \STATE \textbf{Negative item sampling with Gumbel Softmax:} $ y_{\ell}\sim P(y_{\ell})=\frac{\textrm{exp}(\pi_{\theta}(y_{\ell}|S)+G_{y_{\ell}})}{\sum_{y_{\ell}\in \mathcal{I}_{H}}\textrm{exp}(\pi_{\theta}(y_{\ell}|S)+G_{y_{\ell}})},\; \ell=1,...,K$
            \STATE \textbf{Tail item reweighting:} $\omega_i=\frac{\textrm{exp}(\alpha_i)}{\sum_{i=1}^m\textrm{exp}(\alpha_i)} $
            \STATE $\begin{aligned} &\textbf{Parameter update:} \theta \leftarrow \theta+\eta\bigtriangledown_\theta\mathbb{E}_{(S,y_w)\sim\mathcal{D}_m}\\&\omega\mathopen{\Bigg[}\textrm{log}\pi_{\theta}(y_w|S)+\lambda\textrm{log}\\
 &\frac{\textrm{exp}\left(\pi_\theta(y_w|S)/\tau\right)}{\textrm{exp}\left(\pi_\theta(y_w|S)/\tau\right)+\sum_{\ell=1}^K\textrm{exp}\left(\pi_\theta(y_\ell|S)/\tau\right)}\mathclose{\Bigg]}\end{aligned}$
\ENDFOR
\STATE 
\STATE \textsc{OUTPUT:}
\STATE   \textbf{Updated model:} Policy model $\pi_\theta(\cdot)$\;  
\end{algorithmic}
\label{alg1}
\end{algorithm}

\section{Experiments}
\label{sec:exp}

We conduct experiments on three real-world datasets against ten baselines to evaluate the performance of our approach. We aim to answer five research questions:
\begin{itemize}
    \item \textbf{RQ1:} How does \ours compare to state-of-the-art sequential recommendation methods, including representative models, tail-item oriented attempts, and preference alignment solutions, in overall recommendation and tail-item recommendation?
    \item \textbf{RQ2:} What are the training efficiency and GPU memory usage of \ours compared to the state-of-the-art methods?
    \item \textbf{RQ3:} How do key components, \eg negative sampling strategy and reweighting operation, affect the performance of \ours? Can the LPO loss also avail other baselines?
    \item \textbf{RQ4:} How do hyperparameter settings affect the performance of \ours?
    \item \textbf{RQ5:} 
    To what extent does the introduction of the LPO loss affect the recommendation probability distribution of tail items?
\end{itemize}

\subsection{Datasets and Evaluation Metrics}
\label{sec:dataset}

\paratitle{Datasets.} 
We use three subcategories from the Amazon datasets, namely \textit{Amazon Beauty}, \textit{Amazon Toys}, and \textit{Amazon Sports}\footnote{All of them are available at \url{https://nijianmo.github.io/amazon/index.html}.}~\cite{ni2019justifying} for performance evaluation. All these datasets contain rich side information (\eg titles, descriptions, and images associated with items, as well as ratings and reviews from users) and are widely used in various recommendation tasks.
Consistent with prior studies~\cite{kang2018self, bian2023multi, li2025reembedding}, we treat user-item ratings as interaction records while excluding those users and items with fewer than five interactions. 
For each user, the remaining interactions are sorted in chronological order. Then, we employ a leave-one-out approach to partition the data into training, test, and validation sets, \ie the most recent interactions are reserved for testing, the second most recent for validation, and the rest for training.
For instance, given a user sequence $S_i=[x_1,x_2,...,x_{t+1}]$, $x_{t+1}$ is used as the test label, $x_t$ the validation label, and $[x_1,x_2,...,x_{t-1}]$ for model training.
We set the maximum sequence length as $10$ across all datasets, which has proven to be sufficient according to statistical characteristics. As for tail item identification, we apply the Pareto principle~\cite{box1986analysis}, \ie the $20/80$ rules for head and tail item division.

\paratitle{Evaluation Metrics.}
We employ two widely adopted recommendation metrics for performance comparison: HR@N (Hit Rate) and NDCG@N (Normalized Discounted Cumulative Gain).
HR@N reveals the model’s recall capability by quantifying the proportion of relevant items appearing within the top-N recommendations.
NDCG@N measures the model's ranking quality by accounting for the positions of relevant items within the ranked list.
In our experiments, we set $\textrm{N}=5, 10$ and $20$ to facilitate a comparative analysis between our proposed method \ours and baseline models. 
We consider all items as candidates to alleviate the significant discrepancy from real-world scenarios where limited candidates are available~\cite{krichene2022sampled}.

\subsection{Baselines and Implementation Details}
\label{sec:baselines}

\paratitle{Baselines.} 
We select three categories of baselines: 1) sequential recommendation methods, 2) tail-item-focused methods, and 3) preference alignment optimization algorithms for sequential recommendation.
\begin{itemize}
    \item \textbf{Sequential Recommendation Methods.} We select three representative sequential recommendation methods: \textbf{Caser}\footnote{\url{https://github.com/graytowne/caser}}~\cite{tang2018personalized}, \textbf{GRU4Rec}\footnote{\url{https://github.com/hidasib/GRU4Rec}}~\cite{hidasi2015session}, and \textbf{SASRec}\footnote{\url{https://github.com/kang205/SASRec}}~\cite{kang2018self}. These methods represent the three typical neural network architectures used in sequential recommendation: Convolutional Neural Network, GRU, and Transformer.  
    \item \textbf{Tail-item-focused Methods.} We compare our approach with three tail-item-focused sequential recommendation methods to evaluate its effectiveness in improving tail-item recommendation performance: \textbf{CITES}~\cite{jang2020cities}, \textbf{MELT}\footnote{\url{https://github.com/rlqja1107/MELT}}~\cite{kim2023melt}, and \textbf{R2Rec}\footnote{\url{https://github.com/Yuhanleeee/R2Rec}}~\cite{li2025reembedding}.
    CITES and MELT introduce auxiliary information from head items to enhance tail item representations. 
    R2Rec is model-agnostic—it prioritizes tail item optimization during training.
    \item \textbf{Preference Alignment Algorithms.}
    We select four representative preference alignment algorithms closely related to our approach: DPO\footnote{\url{https://github.com/eric-mitchell/direct-preference-optimization}}~\cite{rafailov2023direct}, S-DPO\footnote{\url{https://github.com/chenyuxin1999/S-DPO}}~\cite{chen2024softmax}, SimPO\footnote{\url{https://github.com/princeton-nlp/SimPO}}~\cite{meng2024simpo}, ORPO\footnote{\url{https://github.com/xfactlab/orpo}}~\cite{hong2024reference}.
    DPO explicitly models the reward score under the Bradley-Terry framework~\cite{bradley1952rank} for preference comparison. S-DPO extends DPO to support multi-negative preference comparison based on the full-ranking Plackett-Luce (PL) model. SimPO and ORPO remove the reference model for training efficiency.
\end{itemize}

\paratitle{Implementation Details.}
We employ the Transformer architecture as the backbone for sequential recommendation, setting the block number to $1$ and the multi-head number to $16$.
We further set the embedding and hidden size of Transformer blocks to $768$, consistent with image-based or text-based embedding generated by the CLIP model~\cite{radford2021learning}.
We use the LPO loss (ref. Eq.~\eqref{eq:loss_final}) as the optimization function for model training, apply the Adam optimizer with a learning rate of $5 \times 10^{-4}$, and adopt the warm-up strategy with a step ratio of $0.1$.
With the dropout ratio set to $0.8$ and the batch size to $128$, we fix the hyperparameters in the loss function to $\lambda=0.5$ and $\tau=0.1$ in the overall performance comparison (Section~\ref{comparisonsec}).
We further analyze the hyperparameters of \ours (Section~\ref{sec:hyperparameter}) by varying those two hyperparameters within the ranges: $\lambda\in[0.05, 3]$ and $\tau\in[0.05, 2]$.
For ablation studies (Section~\ref{sec:ablation}), we apply the same model architecture and hyperparameter settings to the four baseline preference alignment algorithms (\ie DPO, S-DPO, SimPO, and ORPO), but let each algorithm use their own losses---instead of the LPO loss in Eq.~\eqref{eq:loss_final}---during training.
The reference models in DPO and S-DPO were obtained from the well-trained optimal SASRec.
We also tune the hyperparameters of the baseline models to their optimal settings to ensure fairness in performance comparison.
All experiments are conducted on Intel Xeon CPU E5-2680 and NVIDIA GeForce RTX 3090 GPU. 

\subsection{Overall and Tail Item Performance (RQ1)}
\label{comparisonsec}

\renewcommand{\arraystretch}{1.24}
\begin{table*}[!t]
\caption{Overall and tail item performance on \textit{AMAZON BEAUTY}, \textit{AMAZON TOYS}, and \textit{AMAZON SPORTS} datasets. The best results are highlighted in boldface, and the second-best results are underlined.
$\blacktriangle\%$ means improvement ($\%$) against the best results.
* denotes a significant improvement over the best baseline results (t-test $P<.05$).\label{tab:overall_tail}}
\resizebox{\textwidth}{!}{
\begin{tabular}{c|clcccccccccccr}
\hline\hline
\multicolumn{2}{c}{\textbf{Dataset}}                                                & \textbf{Metrics} & \textbf{Caser} & \textbf{GRU4Rec} & \textbf{SASRec} & \textbf{MELT} & \textbf{CITES} & \textbf{R2Rec} & \textbf{DPO} & \textbf{S-DPO} & \textbf{SimPO} & \textbf{ORPO} & \textbf{\ours} & $\blacktriangle\%$ \\ \hline\hline
\multirow{18}{*}{\rotatebox[origin=c]{90}{\textbf{Overall Performance}}}  & \multirow{6}{*}{\rotatebox[origin=c]{90}{\textbf{Beauty}}} & \textbf{HR@5}    & 0.0359         & 0.0382           & 0.0523          & 0.0195        & 0.0487         & 0.0518         & 0.0530       & 0.0524         & 0.0521         & \underline{0.0531}        & $\mathbf{0.0542}^*$        & 2.07\%          \\
                                                 &                                  & \textbf{HR@10}   & 0.0527         & 0.0568           & 0.0696          & 0.0380        & 0.0695         & 0.0699         & \underline{0.0713}       & 0.0706         & 0.0694         & 0.0703        & $\mathbf{0.0728}^*$       & 2.10\%         \\
                                                 &                                  & \textbf{HR@20}   & 0.0765         & 0.0821           & 0.0934          & 0.0649        & \underline{0.0963}         & 0.0955         & 0.0957       & 0.0960         & 0.0899         & 0.0959        & $\mathbf{0.0988}^*$        & 2.60\%          \\
                                                 &                                  & \textbf{NDCG@5}  & 0.0258         & 0.0259           & 0.0390          & 0.0114        & 0.0355         & 0.0398         & 0.0405       & \underline{0.0406}         & 0.0400         & 0.0400        & $\mathbf{0.0420}^*$        & 3.45\%          \\
                                                 &                                  & \textbf{NDCG@10} & 0.0311         & 0.0319           & 0.0446          & 0.0173        & 0.0422         & \underline{0.0456}         & 0.0454       & 0.0442         & 0.0452         & 0.0450        & $\mathbf{0.0466}^*$        & 2.19\%          \\
                                                 &                                  & \textbf{NDCG@20} & 0.0371         & 0.0383           & 0.0506          & 0.0241        & 0.0490         & \underline{0.0517}         & 0.0509       & 0.0501         & 0.0503         & \underline{0.0517}        & $\mathbf{0.0531}^*$        & 2.71\%          \\ \cline{2-15} 
                                                 & \multirow{6}{*}{\rotatebox[origin=c]{90}{\textbf{Toys}}}   & \textbf{HR@5}    & 0.0106         & 0.0334           & 0.0546          & 0.0215        & 0.0570         & 0.0577         & 0.0564       & 0.0590         & 0.0594         & \underline{0.0597}        & $\mathbf{0.0608}^*$       & 2.36\%          \\
                                                 &                                  & \textbf{HR@10}   & 0.0156         & 0.0464           & 0.0663          & 0.0379        & 0.0751         & \underline{0.0769}         & 0.0681       & 0.0751         & 0.0757         & 0.0759        & $\mathbf{0.0776}^*$        & 2.24\%          \\
                                                 &                                  & \textbf{HR@20}   & 0.0229         & 0.0645           & 0.0765          & 0.0650        & 0.0940         & 0.0949         & 0.0792       & 0.0907         & 0.0951         & \underline{0.0953}        & $\mathbf{0.0978}^*$        & 2.62\%          \\
                                                 &                                  & \textbf{NDCG@5}  & 0.0075         & 0.0243           & 0.0429          & 0.0122        & 0.0426         & 0.0430         & 0.0439       & \underline{0.0449}         & 0.0444         & 0.0445        & $\mathbf{0.0459}^*$        & 2.23\%          \\
                                                 &                                  & \textbf{NDCG@10} & 0.0091         & 0.0285           & 0.0467          & 0.0174        & 0.0484         & 0.0485         & 0.0477       & 0.0498         & 0.0501         & \underline{0.0502}        & $\mathbf{0.0513}^*$        & 2.19\%          \\
                                                 &                                  & \textbf{NDCG@20} & 0.0109         & 0.0332           & 0.0493          & 0.0242        & 0.0544         & 0.0546         & 0.0506       & 0.0537         & 0.0546         & \underline{0.0548}        & $\mathbf{0.0564}^*$        & 2.92\%          \\ \cline{2-15} 
                                                 & \multirow{6}{*}{\rotatebox[origin=c]{90}{\textbf{Sports}}} & \textbf{HR@5}    & 0.0181         & 0.0195           & 0.0317          & 0.0170        & 0.0278         & \underline{0.0315}         & 0.0306       & 0.0309         & 0.0314         & 0.0312        & $\mathbf{0.0324}^*$        & 2.21\%          \\
                                                 &                                  & \textbf{HR@10}   & 0.0279         & 0.0306           & 0.0433          & 0.0289        & 0.0417         & 0.0435         & 0.0420       & 0.0431         & 0.0426         & \underline{0.0434}        & $\mathbf{0.0447}^*$        & 2.76\%          \\
                                                 &                                  & \textbf{HR@20}   & 0.0431         & 0.0451           & 0.0599          & 0.0477        & \underline{0.0608}         & 0.0604         & 0.0583       & \underline{0.0608}         & 0.0601         & 0.0024        & $\mathbf{0.0623}^*$      & 2.47\%          \\
                                                 &                                  & \textbf{NDCG@5}  & 0.0123         & 0.0132           & \underline{0.0228}          & 0.0103        & 0.0190         & 0.0224         & 0.0222       & 0.0220         & 0.0224         & 0.0219        & $\mathbf{0.0236}^*$        & 3.57\%          \\
                                                 &                                  & \textbf{NDCG@10} & 0.0155         & 0.0168           & 0.0251          & 0.0141        & 0.0235         & 0.0263         & 0.0259       & 0.0259         & \underline{0.0264}         & 0.0261        & $\mathbf{0.0272}^*$        & 3.03\%          \\
                                                 &                                  & \textbf{NDCG@20} & 0.0194         & 0.0204           & 0.0306          & 0.0188        & 0.0283         & 0.0305         & 0.0299       & \underline{0.0308}         & 0.0305         & 0.0304        & $\mathbf{0.0316}^*$       & 2.60\%          \\ \hline\hline
\multirow{18}{*}{\rotatebox[origin=c]{90}{\textbf{Tail Item Performance}}} & \multirow{6}{*}{\rotatebox[origin=c]{90}{\textbf{Beauty}}} & \textbf{HR@5}    & 0.0017         & 0.0116           & 0.0224          & 0.0025        & 0.0043         & \underline{0.0307}         & 0.0187       & 0.0175         & 0.0260         & 0.0206        & $\mathbf{0.0354}^*$        & 15.31\%         \\
                                                 &                                  & \textbf{HR@10}   & 0.0021         & 0.0215           & 0.0265          & 0.0055        & 0.0059         & \underline{0.0335}         & 0.0232       & 0.0208         & 0.0302         & 0.0255        & $\mathbf{0.0373}^*$       & 11.34\%         \\
                                                 &                                  & \textbf{HR@20}   & 0.0031         & 0.0293           & 0.0314          & 0.0076        & 0.0092         & \underline{0.0357}         & 0.0272       & 0.0262         & 0.0331         & 0.0279        & $\mathbf{0.0399}^*$        & 11.76\%         \\
                                                 &                                  & \textbf{NDCG@5}  & 0.0011         & 0.0068           & 0.0150          & 0.0015        & 0.0035         & \underline{0.0234}         & 0.0154       & 0.0146         & 0.0198         & 0.0160        & $\mathbf{0.0307}^*$       & 31.20\%         \\
                                                 &                                  & \textbf{NDCG@10} & 0.0013         & 0.0101           & 0.0163          & 0.0025        & 0.0040         & \underline{0.0243}         & 0.0169       & 0.0157         & 0.0212         & 0.0175        & $\mathbf{0.0313}^*$       & 28.81\%         \\
                                                 &                                  & \textbf{NDCG@20} & 0.0015         & 0.0121           & 0.0175          & 0.0030        & 0.0049         & \underline{0.0249}         & 0.0179       & 0.0171         & 0.0219         & 0.0181        & $\mathbf{0.0319}^*$      & 28.11\%         \\ \cline{2-15} 
                                                 & \multirow{6}{*}{\rotatebox[origin=c]{90}{\textbf{Toys}}} & \textbf{HR@20}  & 0.0009         & 0.0202           & 0.0474          & 0.0014        & 0.0247         & \underline{0.0494}         & 0.0433       & 0.0417         & 0.0424         & 0.0417        & $\mathbf{0.0505}^*$      & 2.23\%          \\
                                                 &                                  & \textbf{HR@10}   & 0.0011         & 0.0271           & 0.0472          & 0.0021        & 0.0312         & \underline{0.0521}         & 0.0468       & 0.0482         & 0.0485         & 0.0485        & $\mathbf{0.0534}^*$      & 2.50\%          \\
                                                 &                                  & \textbf{HR@20}   & 0.0018         & 0.0368           & 0.0532          & 0.0108        & 0.0415         & 0.0529         & 0.0518       & 0.0525         & \underline{0.0534}         & 0.0527        & $\mathbf{0.0550}^*$       & 3.38\%          \\
                                                 &                                  & \textbf{NDCG@5}  & 0.0005         & 0.0154           & 0.0360         & 0.0006        & 0.0184         & \underline{0.0390}         & 0.0351       & 0.0284         & 0.0283         & 0.0277        & $\mathbf{0.0419}^*$      & 7.44\%          \\
                                                 &                                  & \textbf{NDCG@10} & 0.0006         & 0.0177           & 0.0379          & 0.0008        & 0.0205         & 0.0380         & 0.0370       & 0.0306         & 0.0303         & \underline{0.0399}        & $\mathbf{0.0429}^*$       & 7.52\%          \\
                                                 &                                  & \textbf{NDCG@20} & 0.0008         & 0.0202           & 0.0395          & 0.0029        & 0.0231         & \underline{0.0401}         & 0.0376       & 0.0316         & 0.0315         & 0.0310        & $\mathbf{0.0433}^*$        & 7.98\%          \\ \cline{2-15} 
                                                 & \multirow{6}{*}{\rotatebox[origin=c]{90}{\textbf{Sports}}} & \textbf{HR@5}    & 0.0005         & 0.0066           & 0.0020          & 0.0006        & 0.0008         & \underline{0.0047}         & 0.0041       & 0.0035         & 0.0035         & 0.0028        & $\mathbf{0.0054}^*$       & 31.70\%          \\
                                                 &                                  & \textbf{HR@10}   & 0.0008         & 0.0043           & 0.0028          & 0.0006        & 0.0018         & \underline{0.0054}         & 0.0049       & 0.0043         & 0.0041         & 0.0046        & $\mathbf{0.0069}^*$       & 27.78\%         \\
                                                 &                                  & \textbf{HR@20}   & 0.0012         & 0.0119           & 0.0046          & 0.0011        & 0.0023         & \underline{0.0064}         & 0.0063       & 0.0050         & 0.0050         & 0.0054        & $\mathbf{0.0096}^*$       & 50.00\%            \\
                                                 &                                  & \textbf{NDCG@5}  & 0.0003         & 0.0025           & 0.0015          & 0.0003        & 0.0008         & \underline{0.0031}         & 0.0026       & 0.0026         & 0.0021         & 0.0018        & $\mathbf{0.0040}^*$        & 29.03\%         \\
                                                 &                                  & \textbf{NDCG@10} & 0.0004         & 0.0044           & 0.0017          & 0.0003        & 0.0011         & \underline{0.0032}         & 0.0029       & 0.0028         & 0.0023         & 0.0024        & $\mathbf{0.0045}^*$       & 40.63\%         \\
                                                 &                                  & \textbf{NDCG@20} & 0.0005         & 0.0031           & 0.0022          & 0.0004        & 0.0012         & \underline{0.0035}         & 0.0032       & 0.0030         & 0.0026         & 0.0026        & $\mathbf{0.0052}^*$       & 48.57\%         \\ \hline\hline
\end{tabular}}
\end{table*}

\paratitle{Overall Performance.} The overall performance of \ours in comparison with baselines is shown in Table~\ref{tab:overall_tail} (upper part). The results show that 1) SASRec is a competitive baseline model against other representative solutions on all the three datasets; 2) incorporating the auxiliary preference alignment losses (\eg DPO, S-DPO, SimPO, and ORPO) consistently improves the performance of SASRec, validating the effectiveness of preference alignment optimization in general recommendation tasks; 3) \ours achieves the optimal results on all the metrics across the three datasets, showing superiority to other baselines.

\paratitle{Tail Item Performance.} Performance of the compared methods in tail-item recommendation is shown in Table~\ref{tab:overall_tail} (lower part).
Our approach, \ours, consistently outperforms the baselines across all metrics.
In particular, \ours outperforms all the baselines by a large margin on \textit{Amazon Beauty} and \textit{Amazon Sports} datasets---it achieves up to $50\%$ and $48.57\%$ improvement in the HR and NDCG metrics, respectively, demonstrating the superiority of our LPO loss in enhancing tail-item recommendation.
R2Rec achieves suboptimal results on $15$ out of all $18$ metrics, thanks to its reweighting strategy, which contributes to allocating more attention to tail items during model optimization.
Preference alignment solutions are also demonstrated to enhance the recommendation of tail items when compared with representative sequential recommendation methods.
All methods demonstrate comparable performance on \textit{Amazon Toys} dataset---the much lower number of historical interaction records than those of \textit{Amazon Beauty} and \textit{Amazon Sports} (\ie $121,285$ vs. $207,900$) causes a significant challenge for capturing complex patterns for tail items.

\subsection{Training Efficiency (RQ2)}
\label{sec:efficiency}

\begin{figure}[!t]
\centering
\includegraphics[width=\columnwidth]{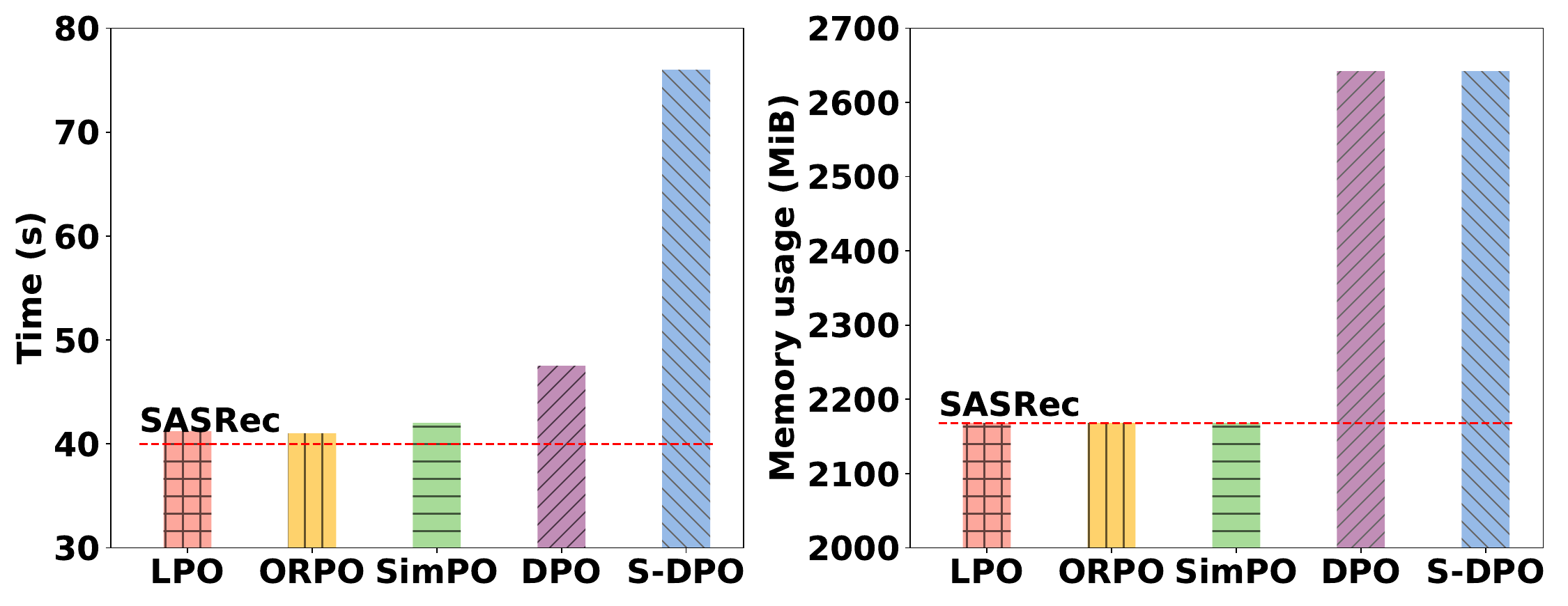}
\caption{The single-pass (one epoch) training time and peak GPU memory usage of our \ours against other preference alignment baselines. All the experiments are conducted on the \textit{Amazon Beauty} dataset under the same hyperparameter settings and configurations. Each experiment is repeated five times, and the average values are reported as the final results. Our method achieves comparable performance to SASRec (red dot line) while incurring significantly lower computational cost than DPO and S-DPO.}
\label{fig:efficiency}
\end{figure}

We investigate the training efficiency of our LPO loss against other preference alignment solutions (DPO, S-DPO, SimPO, and ORPO) by reporting the single-pass training time and peak GPU memory usage on the \textit{Amazon Beauty} dataset in Fig.~\ref{fig:efficiency}.
We apply the same hyperparameter settings and configurations to all the methods to ensure a fair comparison.
Then, we repeat each experiment five times and record the average of five runs as the final performance.
\ours maintains the low training time and memory usage of vanilla SASRec and remains comparable to ORPO and SimPO.
It demonstrates the superior efficiency of the LPO loss against DPO and S-DPO, both of which involve a reference model that incurs additional computational cost during training.

\subsection{Ablation Studies (RQ3)}
\label{sec:ablation}

\begin{table*}[]
\caption{Overall and tail item performance on the \textit{AMAZON BEAUTY} datasets. \textit{txt} and \textit{img} denote image embedding and text embedding generated by CLIP based on item title and image information, respectively.The best results are highlighted in boldface, and the second-best results are underlined.\label{tab:ablation}}
\begin{center}
\begin{tabular}{lccccccccc}
\hline\hline
\multirow{2}{*}{\textbf{Methods}} & \multicolumn{4}{c}{\textbf{Overall Performance}}                    & \textbf{} & \multicolumn{4}{c}{\textbf{Tail Item Performance}}                  \\ \cline{2-5} \cline{7-10} 
                                  & \textbf{HR@5} & \textbf{HR@10} & \textbf{NDCG@5} & \textbf{NDCG@10} & \textbf{} & \textbf{HR@5} & \textbf{HR@10} & \textbf{NDCG@5} & \textbf{NDCG@10} \\ \hline\hline
\textbf{LPO4Rec}                  & 0.0542        & 0.0728         & \textbf{0.0420}          & 0.0466           &           & \textbf{0.0354}        & \underline{0.0373}         & \textbf{0.0307}          & \textbf{0.0313}           \\
\textbf{LPO4Rec w. img}           & \textbf{0.0550}        & \textbf{0.0768}         & 0.0405          & \textbf{0.0472}           &           & \underline{0.0347}        & \textbf{0.0461}         & \underline{0.0237}          & \underline{0.0272}           \\
\textbf{LPO4Rec w. txt}           & \underline{0.0544}        & \underline{0.0755}         & 0.0396          & \underline{0.0468}           &           & 0.0113        & 0.0151         & 0.0090          & 0.0102           \\ \hline
\textbf{LPO4Rec w/o LPO loss}          & 0.0523        & 0.0696         & 0.0390          & 0.0446           &           & 0.0224        & 0.0265         & 0.0150          & 0.0163           \\
\textbf{LPO4Rec w/o reweight}     & 0.0531        & 0.0706         & 0.0391          & 0.0448           &           & 0.0227        & 0.0283         & 0.0149          & 0.0168           \\
\textbf{LPO4Rec w. Top-K selection}          & 0.0541        & 0.0721         & \underline{0.0406}          & 0.0464           &           & 0.0302        & 0.0331         & 0.0235          & 0.0245           \\
\textbf{LPO4Rec w. random sampling}        & 0.0534        & 0.0709         & 0.0403          & 0.0460           &           & 0.0291        & 0.0335         & 0.0232          & 0.0247           \\ \hline
\textbf{GRU4Rec}                      & 0.0382        & 0.0568         & 0.0259          & 0.0319           &           & 0.0116        & 0.0215         & 0.0068          & 0.0101           \\
\textbf{GRU4Rec w. LPO loss \& reweight}   & 0.0456        & 0.0541         & 0.0287          & 0.0304           &           & 0.0243        & 0.0291         & 0.0178          & 0.0193           \\
\textbf{Caser}                    & 0.0359        & 0.0527         & 0.0258          & 0.0311           &           & 0.0017        & 0.0021         & 0.0011          & 0.0013           \\
\textbf{Caser w. LPO loss \& reweight} & 0.0418        & 0.0589         & 0.0293          & 0.0348           &           & 0.0083        & 0.0113         & 0.0067          & 0.0074           \\ \hline\hline
\end{tabular}
\end{center}
\end{table*}

We evaluate the effectiveness of individual modules by comparing the performance before and after removing a module from our overall approach on the \textit{Amazon Beauty} dataset. 
Additionally, we incorporate modules of our approach into GRU4Rec and Caser to evaluate their impact across different model architectures.
LLM-driven solutions to sequential recommendation generally concatenate associated text information of historical items as a sequence and then feed it into the LLMs for next-item generation.
As such, they often become intractable, given large historical interactions~\cite{hu2024enhancing}.
Therefore, we derive a variant of \ours by employing LLMs/LVMs to encode the item-associated text/image information---which replaces ID-based embedding---instead of following the LLM-driven paradigm.

Specifically, we compare the following variants:
\begin{itemize}
    \item \textbf{w. img or w. txt:} obtaining item images or titles with a maximum of $10$ tokens, then applying the image or text encoder from CLIP\footnote{\url{https://huggingface.co/openai/clip-vit-base-patch32}}~\cite{radford2021learning}, respectively, to obtain image-based or text-based item embedding, which replaces ID-based embedding in \ours.
    \item \textbf{w/o LPO loss or w/o reweight}: removing the LPO loss or reweighting strategy from Eq.~\eqref{eq:loss_final}.
    \item \textrm{w. Top-K selection or random sampling}: replacing the Gumbel-Top-K sampling strategy in Eq.~\eqref{eq:neg_sample_gumbel} with the Top-K selection strategy (\ie select K head items with the highest predicted probabilities) or the random sampling strategy (\ie random sample K head items).
    \item \textbf{w. LPO loss \& reweight}: applying Eq.~\eqref{eq:loss_final} as the loss function for SASRec and Caser optimization.
\end{itemize}

Our ablation results (Table~\ref{tab:ablation}) show that incorporating the LPO loss and our reweighting strategy consistently outperforms variants without them, and removing any tailored module or replacing our sampling strategy with any alternative one deteriorates the performance.
Similar conclusions are drawn from the other two datasets.
Although replacing ID-based item embedding with image- or text-based embedding improves the overall performance, it compromises the performance on tail items.
Such a trade-off lies in the \textit{knowledge transfer tax}~\cite{li2025reembedding}, \ie the external knowledge encapsulated in text-based or image-based embedding interferes with the model's optimization directions when the embeddings lack sufficient training, ultimately hurting the recommendation performance. This phenomenon aggravates for tail items, as they suffer from data sparsity relative to head items during training.

\begin{figure*}[!t]
\centering
\includegraphics[width=\textwidth]{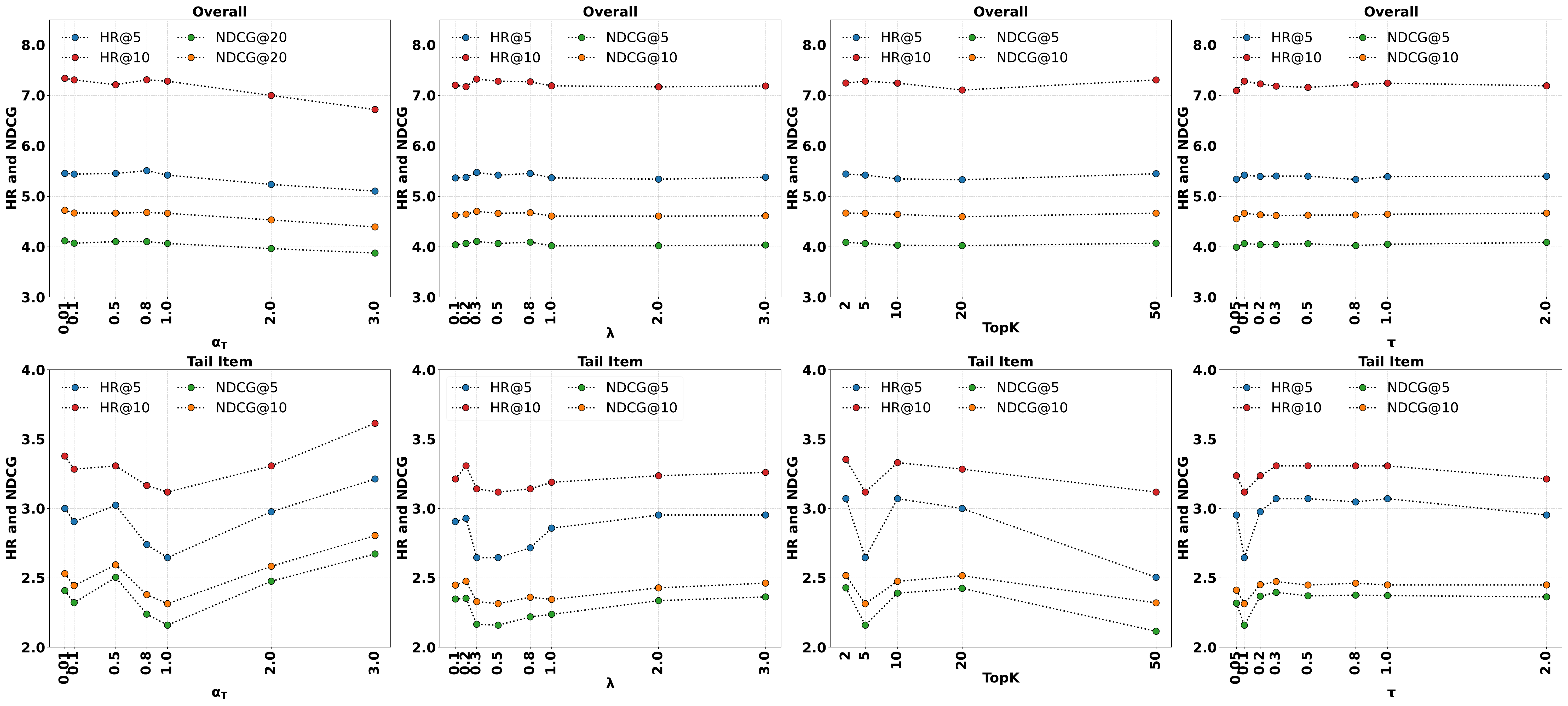}
\caption{The overall and tail item performance under different hyperparameter settings. $\alpha_T$ controls the weight (\ie attention) toward the tail items during model optimization. $\lambda$ balances the cross-entropy loss and LPO loss, governing the trade-off between recommendation optimization and tail item preference alignment optimization. Top-K and $\tau$ jointly control the negative sampling distribution for enhancing preference alignment of tail items.}
\label{fig:hyper}
\end{figure*}

\subsection{Impact of Hyperparameters (RQ4)}
\label{sec:hyperparameter}

We analyze the impact of hyperparameters by evaluating the overall and tail-item recommendation performance across varying settings of $\alpha_T, \lambda$, Top-K, and $\tau$.
Specifically, we varied the model’s attention toward tail items, \ie $\alpha_T$ in Eq.~\eqref{eq:reweight}, within the range of $[0.01, 3.0]$, where a larger $\alpha_T$ encourages the model to focus more on optimizing tail-item performance during training.
We explore the effect of $\lambda$, which balances the trade-off between the cross-entropy loss and the LPO loss in Eq.~\eqref{eq:loss_final}, by increasing its value from $0.1$ to $3.0$, where a larger $\lambda$ places greater emphasis on tail item preference alignment optimization.
Since Top-K and $\tau$ in Eq.~\eqref{eq:reweight} jointly govern the negative sampling strategy by adjusting the sampling distribution, we tune Top-K values within the range $[2, 50]$ and $\tau$ within $[0.05, 2.0]$ to investigate their impact on the final performance.
Our results (Fig.~\ref{fig:hyper}) reveal that 1) increasing $\alpha_T$ steadily enhances tail item performance, with a trade-off in overall performance, which is consistent with our expectations; 2) varying $\lambda$ incurs modest fluctuations in both overall and tail-item performance---while a larger $\lambda$ nudges tail item performance at the cost of slightly decreasing overall performance, we recommend a smaller $\lambda$ (\eg $\lambda=0.3$); 3) Top-K and $\tau$ present opposing effects on overall and tail item performance----while a larger Top-K improves overall recommendation, the tail item performance degrades as a side effect; $\tau$ has an opposite effect.
On balance, we recommend a moderate Top-K value (\eg $10$) to safeguard efficiency.

\subsection{Case Studies (RQ5)}
\label{sec:cases}

\begin{figure}[!t]
\centering
\includegraphics[width=0.9\columnwidth]{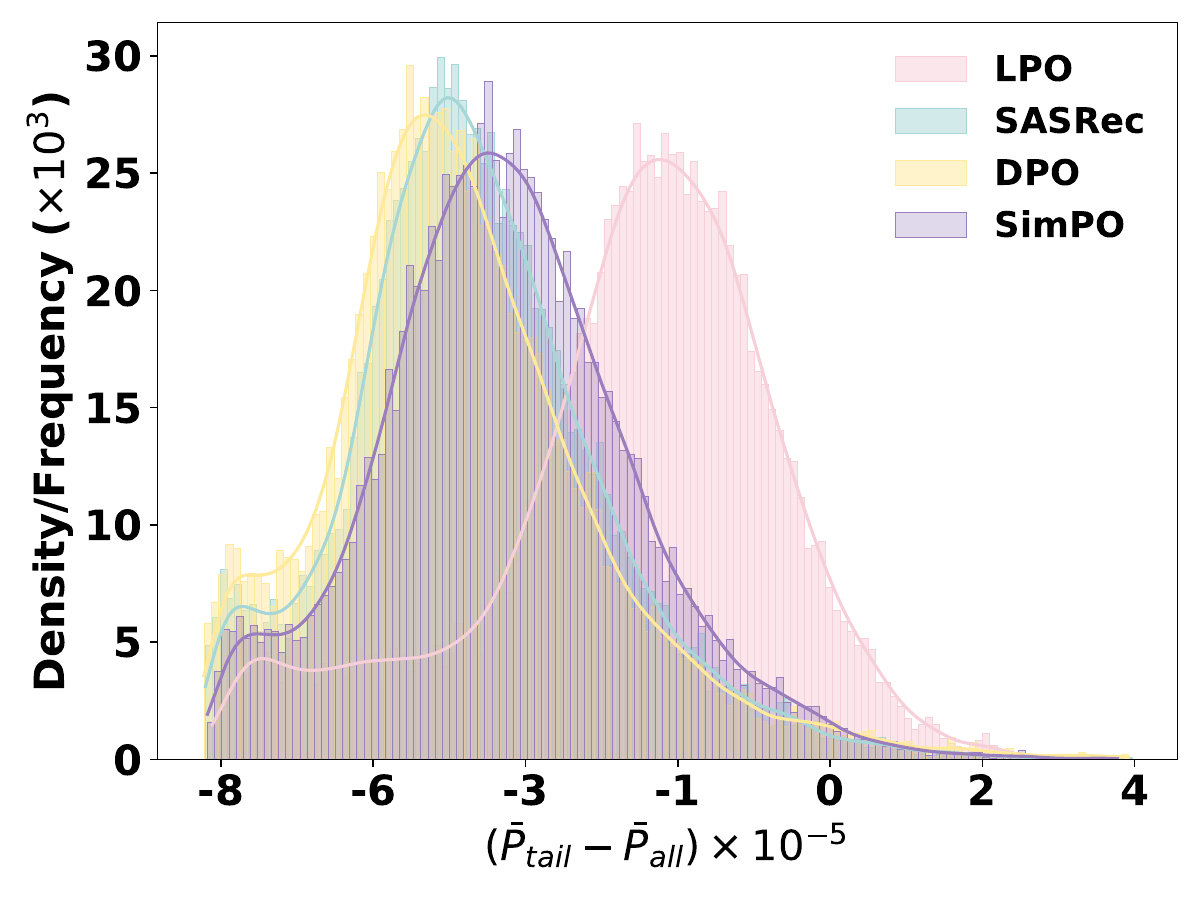}
\caption{The frequency distribution of the differences between the average recommendation probabilities for tail items and all items across all users in \textit{Amazon Beauty} dataset, \ie $\bar{P}_{tail}-\bar{P}_{all}$. The more the distribution is skewed to the right, the higher the recommendation probabilities of tail items are.}
\label{fig:case}
\end{figure}

We plot the frequency distribution of the differences between the average recommendation probabilities of tail items and all items across all users of the \textit{Amazon Beauty} dataset in Fig.~\ref{fig:case}.
The plot shows that incorporating the LPO loss brings about an evident shift of the distribution towards tail items, with the peak value drawing closer to 0.
Incorporating the LPO loss, \ours achieves higher recommendation probabilities on tail items compared to vanilla SASRec and other preference alignment approaches like DPO and SimPO, demonstrating its effectiveness in improving tail-item recommendation performance.

\section{Conclusion}
\label{sec:conclusion}

Tail-item recommendation is a persistent challenge in recommendation research.
We have introduced preference alignment optimization to tail-item recommendation, given its success in LLM applications.
We first extend the pairwise comparison in the Bradley-Terry model to a listwise framework, allowing more efficient exploitation of negative samples during training.
We further propose an adaptive negative sampling strategy with the Gumbel-Softmax trick and a reweighting strategy to refocus the model on tail item optimization, enhancing its performance in tail-item recommendation.
Our proposed \ours achieves up to $50.0\%$ improvement on tail-item recommendation while consuming only $54.2\%$ single-pass training time and $82.1\%$ GPU memory when compared with the state-of-the-art preference alignment method, S-DPO.
Our theoretical analysis further confirms the merits of \ours in hard-item and tail-item optimization.




\bibliographystyle{IEEEtran}
\bibliography{IEEEabrv,ref}

\begin{thebibliography}{10}
\providecommand{\url}[1]{#1}
\csname url@samestyle\endcsname
\providecommand{\newblock}{\relax}
\providecommand{\bibinfo}[2]{#2}
\providecommand{\BIBentrySTDinterwordspacing}{\spaceskip=0pt\relax}
\providecommand{\BIBentryALTinterwordstretchfactor}{4}
\providecommand{\BIBentryALTinterwordspacing}{\spaceskip=\fontdimen2\font plus
\BIBentryALTinterwordstretchfactor\fontdimen3\font minus \fontdimen4\font\relax}
\providecommand{\BIBforeignlanguage}[2]{{%
\expandafter\ifx\csname l@#1\endcsname\relax
\typeout{** WARNING: IEEEtran.bst: No hyphenation pattern has been}%
\typeout{** loaded for the language `#1'. Using the pattern for}%
\typeout{** the default language instead.}%
\else
\language=\csname l@#1\endcsname
\fi
#2}}
\providecommand{\BIBdecl}{\relax}
\BIBdecl

\bibitem{yang2024harnessing}
J.~Yang, H.~Jin, R.~Tang, X.~Han, Q.~Feng, H.~Jiang, S.~Zhong, B.~Yin, and X.~Hu, ``Harnessing the power of llms in practice: A survey on chatgpt and beyond,'' \emph{ACM Transactions on Knowledge Discovery from Data}, vol.~18, no.~6, pp. 1--32, 2024.

\bibitem{zhao2024recommender}
Z.~Zhao, W.~Fan, J.~Li, Y.~Liu, X.~Mei, Y.~Wang, Z.~Wen, F.~Wang, X.~Zhao, J.~Tang, and Q.~Li, ``Recommender systems in the era of large language models (llms),'' \emph{IEEE Transactions on Knowledge and Data Engineering}, vol.~36, no.~11, pp. 6889--6907, 2024.

\bibitem{bao2023tallrec}
K.~Bao, J.~Zhang, Y.~Zhang, W.~Wang, F.~Feng, and X.~He, ``Tallrec: An effective and efficient tuning framework to align large language model with recommendation,'' in \emph{Proceedings of the 17th ACM Conference on Recommender Systems}, 2023, pp. 1007--1014.

\bibitem{bai2024aligning}
Z.~Bai, N.~Wu, F.~Cai, X.~Zhu, and Y.~Xiong, ``Aligning large language model with direct multi-preference optimization for recommendation,'' in \emph{Proceedings of the 33rd ACM International Conference on Information and Knowledge Management}, 2024, pp. 76--86.

\bibitem{li2024align}
Y.~Li, Q.~Zhao, C.~Lin, J.~Su, and Z.~Zhang, ``Who to align with: Feedback-oriented multi-modal alignment in recommendation systems,'' in \emph{Proceedings of the 47th International ACM SIGIR Conference on Research and Development in Information Retrieval}, 2024, pp. 667--676.

\bibitem{deng2025onerec}
J.~Deng, S.~Wang, K.~Cai, L.~Ren, Q.~Hu, W.~Ding, Q.~Luo, and G.~Zhou, ``Onerec: Unifying retrieve and rank with generative recommender and iterative preference alignment,'' \emph{arXiv preprint arXiv:2502.18965}, 2025.

\bibitem{jiang2024survey}
R.~Jiang, K.~Chen, X.~Bai, Z.~He, J.~Li, M.~Yang, T.~Zhao, L.~Nie, and M.~Zhang, ``A survey on human preference learning for large language models,'' \emph{arXiv preprint arXiv:2406.11191}, 2024.

\bibitem{ouyang2022training}
L.~Ouyang, J.~Wu, X.~Jiang, D.~Almeida, C.~Wainwright, P.~Mishkin, C.~Zhang, S.~Agarwal, K.~Slama, A.~Ray \emph{et~al.}, ``Training language models to follow instructions with human feedback,'' \emph{Advances in Neural Information Processing Systems}, vol.~35, pp. 27\,730--27\,744, 2022.

\bibitem{harte2023leveraging}
J.~Harte, W.~Zorgdrager, P.~Louridas, A.~Katsifodimos, D.~Jannach, and M.~Fragkoulis, ``Leveraging large language models for sequential recommendation,'' in \emph{Proceedings of the 17th ACM Conference on Recommender Systems}, 2023, pp. 1096--1102.

\bibitem{zheng2024harnessing}
Z.~Zheng, W.~Chao, Z.~Qiu, H.~Zhu, and H.~Xiong, ``Harnessing large language models for text-rich sequential recommendation,'' in \emph{Proceedings of the ACM Web Conference}, 2024, pp. 3207--3216.

\bibitem{schulman2017proximal}
J.~Schulman, F.~Wolski, P.~Dhariwal, A.~Radford, and O.~Klimov, ``Proximal policy optimization algorithms,'' \emph{arXiv preprint arXiv:1707.06347}, 2017.

\bibitem{shao2024deepseekmath}
Z.~Shao, P.~Wang, Q.~Zhu, R.~Xu, J.~Song, X.~Bi, H.~Zhang, M.~Zhang, Y.~Li, Y.~Wu \emph{et~al.}, ``Deepseekmath: Pushing the limits of mathematical reasoning in open language models,'' \emph{arXiv preprint arXiv:2402.03300}, 2024.

\bibitem{yu2025dapo}
Q.~Yu, Z.~Zhang, R.~Zhu, Y.~Yuan, X.~Zuo, Y.~Yue, T.~Fan, G.~Liu, L.~Liu, X.~Liu \emph{et~al.}, ``Dapo: An open-source llm reinforcement learning system at scale,'' \emph{arXiv preprint arXiv:2503.14476}, 2025.

\bibitem{rafailov2023direct}
R.~Rafailov, A.~Sharma, E.~Mitchell, C.~D. Manning, S.~Ermon, and C.~Finn, ``Direct preference optimization: Your language model is secretly a reward model,'' \emph{Advances in Neural Information Processing Systems}, vol.~36, pp. 53\,728--53\,741, 2023.

\bibitem{meng2024simpo}
Y.~Meng, M.~Xia, and D.~Chen, ``Simpo: Simple preference optimization with a reference-free reward,'' \emph{Advances in Neural Information Processing Systems}, vol.~37, pp. 124\,198--124\,235, 2024.

\bibitem{hong2024reference}
J.~Hong, N.~Lee, and J.~Thorne, ``Reference-free monolithic preference optimization with odds ratio,'' \emph{arXiv preprint arXiv:2403.07691v1}, 2024.

\bibitem{bradley1952rank}
R.~A. Bradley and M.~E. Terry, ``Rank analysis of incomplete block designs: I. the method of paired comparisons,'' \emph{Biometrika}, vol.~39, no. 3/4, pp. 324--345, 1952.

\bibitem{chen2024softmax}
Y.~Chen, J.~Tan, A.~Zhang, Z.~Yang, L.~Sheng, E.~Zhang, X.~Wang, and T.-S. Chua, ``On softmax direct preference optimization for recommendation,'' \emph{arXiv preprint arXiv:2406.09215}, 2024.

\bibitem{shi2023theories}
W.~Shi, J.~Chen, F.~Feng, J.~Zhang, J.~Wu, C.~Gao, and X.~He, ``On the theories behind hard negative sampling for recommendation,'' in \emph{Proceedings of the ACM Web Conference}, 2023, pp. 812--822.

\bibitem{ma2024negative}
H.~Ma, R.~Xie, L.~Meng, F.~Feng, X.~Du, X.~Sun, Z.~Kang, and X.~Meng, ``Negative sampling in recommendation: A survey and future directions,'' \emph{arXiv preprint arXiv:2409.07237}, 2024.

\bibitem{box1986analysis}
G.~E. Box and R.~D. Meyer, ``An analysis for unreplicated fractional factorials,'' \emph{Technometrics}, vol.~28, no.~1, pp. 11--18, 1986.

\bibitem{ni2019justifying}
J.~Ni, J.~Li, and J.~McAuley, ``Justifying recommendations using distantly-labeled reviews and fine-grained aspects,'' in \emph{Proceedings of the 2019 Conference on Empirical Methods in Natural Language Processing and the 9th International Joint Conference on Natural Language Processing}, 2019, pp. 188--197.

\bibitem{shani2005mdp}
G.~Shani, D.~Heckerman, and R.~I. Brafman, ``An mdp-based recommender system,'' \emph{Journal of Machine Learning Research}, vol.~6, no. Sep, pp. 1265--1295, 2005.

\bibitem{rendle2010factorizing}
S.~Rendle, C.~Freudenthaler, and L.~Schmidt-Thieme, ``Factorizing personalized markov chains for next-basket recommendation,'' in \emph{Proceedings of the 19th International Conference on World Wide Web}, 2010, pp. 811--820.

\bibitem{zhang2019deep}
S.~Zhang, L.~Yao, A.~Sun, and Y.~Tay, ``Deep learning based recommender system: A survey and new perspectives,'' \emph{ACM Computing Surveys}, vol.~52, no.~1, pp. 1--38, 2019.

\bibitem{li2024graph}
Z.~Li, C.~Yang, Y.~Chen, X.~Wang, H.~Chen, G.~Xu, L.~Yao, and M.~Sheng, ``Graph and sequential neural networks in session-based recommendation: A survey,'' \emph{ACM Computing Surveys}, vol.~57, no.~2, pp. 1--37, 2024.

\bibitem{pouyanfar2018survey}
S.~Pouyanfar, S.~Sadiq, Y.~Yan, H.~Tian, Y.~Tao, M.~P. Reyes, M.-L. Shyu, S.-C. Chen, and S.~S. Iyengar, ``A survey on deep learning: Algorithms, techniques, and applications,'' \emph{ACM Computing Surveys}, vol.~51, no.~5, pp. 1--36, 2018.

\bibitem{lecun2015deep}
Y.~LeCun, Y.~Bengio, and G.~Hinton, ``Deep learning,'' \emph{Nature}, vol. 521, no. 7553, pp. 436--444, 2015.

\bibitem{yang2025deep}
C.~Yang, Y.~Chen, Z.~Li, X.~Wang, K.~Shi, L.~Yao, G.~Xu, and Z.~Guo, ``Deep multimodal learning for time series analysis in social computing: a survey,'' \emph{International Journal of Multimedia Information Retrieval}, vol.~14, no.~2, p.~15, 2025.

\bibitem{tang2018personalized}
J.~Tang and K.~Wang, ``Personalized top-n sequential recommendation via convolutional sequence embedding,'' in \emph{Proceedings of the 11th ACM International Conference on Web Search and Data Mining}, 2018, pp. 565--573.

\bibitem{hidasi2015session}
B.~Hidasi, A.~Karatzoglou, L.~Baltrunas, and D.~Tikk, ``Session-based recommendations with recurrent neural networks,'' \emph{arXiv preprint arXiv:1511.06939}, 2015.

\bibitem{xie2021adversarial}
Z.~Xie, C.~Liu, Y.~Zhang, H.~Lu, D.~Wang, and Y.~Ding, ``Adversarial and contrastive variational autoencoder for sequential recommendation,'' in \emph{Proceedings of the ACM Web Conference}, 2021, pp. 449--459.

\bibitem{kang2018self}
W.-C. Kang and J.~McAuley, ``Self-attentive sequential recommendation,'' in \emph{Proceedings of the 18th IEEE International Conference on Data Mining}.\hskip 1em plus 0.5em minus 0.4em\relax IEEE, 2018, pp. 197--206.

\bibitem{sun2019bert4rec}
F.~Sun, J.~Liu, J.~Wu, C.~Pei, X.~Lin, W.~Ou, and P.~Jiang, ``Bert4rec: Sequential recommendation with bidirectional encoder representations from transformer,'' in \emph{Proceedings of the 28th ACM International Conference on Information and Knowledge Management}, 2019, pp. 1441--1450.

\bibitem{fan2021continuous}
Z.~Fan, Z.~Liu, J.~Zhang, Y.~Xiong, L.~Zheng, and P.~S. Yu, ``Continuous-time sequential recommendation with temporal graph collaborative transformer,'' in \emph{Proceedings of the 30th ACM International Conference on Information and Knowledge Management}, 2021, pp. 433--442.

\bibitem{wu2020sse}
L.~Wu, S.~Li, C.-J. Hsieh, and J.~Sharpnack, ``Sse-pt: Sequential recommendation via personalized transformer,'' in \emph{Proceedings of the 14th ACM Conference on Recommender Systems}, 2020, pp. 328--337.

\bibitem{xia2022multi}
L.~Xia, C.~Huang, Y.~Xu, and J.~Pei, ``Multi-behavior sequential recommendation with temporal graph transformer,'' \emph{IEEE Transactions on Knowledge and Data Engineering}, vol.~35, no.~6, pp. 6099--6112, 2022.

\bibitem{wang2020kerl}
P.~Wang, Y.~Fan, L.~Xia, W.~X. Zhao, S.~Niu, and J.~Huang, ``Kerl: A knowledge-guided reinforcement learning model for sequential recommendation,'' in \emph{Proceedings of the 43rd International ACM SIGIR conference on research and development in Information Retrieval}, 2020, pp. 209--218.

\bibitem{li2023diffurec}
Z.~Li, A.~Sun, and C.~Li, ``Diffurec: A diffusion model for sequential recommendation,'' \emph{ACM Transactions on Information Systems}, vol.~42, no.~3, pp. 1--28, 2023.

\bibitem{wang2023diffusion}
W.~Wang, Y.~Xu, F.~Feng, X.~Lin, X.~He, and T.-S. Chua, ``Diffusion recommender model,'' in \emph{Proceedings of the 46th International ACM SIGIR Conference on Research and Development in Information Retrieval}, 2023, pp. 832--841.

\bibitem{hou2022towards}
Y.~Hou, S.~Mu, W.~X. Zhao, Y.~Li, B.~Ding, and J.-R. Wen, ``Towards universal sequence representation learning for recommender systems,'' in \emph{Proceedings of the 28th ACM SIGKDD Conference on Knowledge Discovery and Data Mining}, 2022, pp. 585--593.

\bibitem{li2023text}
J.~Li, M.~Wang, J.~Li, J.~Fu, X.~Shen, J.~Shang, and J.~McAuley, ``Text is all you need: Learning language representations for sequential recommendation,'' in \emph{Proceedings of the 29th ACM SIGKDD Conference on Knowledge Discovery and Data Mining}, 2023, pp. 1258--1267.

\bibitem{geng2022recommendation}
S.~Geng, S.~Liu, Z.~Fu, Y.~Ge, and Y.~Zhang, ``Recommendation as language processing (rlp): A unified pretrain, personalized prompt \& predict paradigm (p5),'' in \emph{Proceedings of the 16th ACM Conference on Recommender Systems}, 2022, pp. 299--315.

\bibitem{shu2024rah}
Y.~Shu, H.~Zhang, H.~Gu, P.~Zhang, T.~Lu, D.~Li, and N.~Gu, ``Rah! recsys--assistant--human: A human-centered recommendation framework with llm agents,'' \emph{IEEE Transactions on Computational Social Systems}, 2024.

\bibitem{peng2023towards}
B.~Peng, B.~Burns, Z.~Chen, S.~Parthasarathy, and X.~Ning, ``Towards efficient and effective adaptation of large language models for sequential recommendation,'' \emph{arXiv preprint arXiv:2310.01612}, 2023.

\bibitem{zhao2024raserec}
X.~Zhao, B.~Hu, Y.~Zhong, S.~Huang, Z.~Zheng, M.~Wang, H.~Wang, and M.~Zhang, ``Raserec: Retrieval-augmented sequential recommendation,'' \emph{arXiv preprint arXiv:2412.18378}, 2024.

\bibitem{ye2025harnessing}
Y.~Ye, Z.~Zheng, Y.~Shen, T.~Wang, H.~Zhang, P.~Zhu, R.~Yu, K.~Zhang, and H.~Xiong, ``Harnessing multimodal large language models for multimodal sequential recommendation,'' in \emph{Proceedings of the AAAI Conference on Artificial Intelligence}, vol.~39, no.~12, 2025, pp. 13\,069--13\,077.

\bibitem{yin2012challenging}
H.~Yin, B.~Cui, J.~Li, J.~Yao, and C.~Chen, ``Challenging the long tail recommendation,'' \emph{Proceedings of the VLDB Endowment}, vol.~5, no.~9, 2012.

\bibitem{li2017two}
J.~Li, K.~Lu, Z.~Huang, and H.~T. Shen, ``Two birds one stone: on both cold-start and long-tail recommendation,'' in \emph{Proceedings of the 25th ACM International Conference on Multimedia}, 2017, pp. 898--906.

\bibitem{huang2006correcting}
J.~Huang, A.~Gretton, K.~Borgwardt, B.~Sch{\"o}lkopf, and A.~Smola, ``Correcting sample selection bias by unlabeled data,'' \emph{Advances in Neural Information Processing Systems}, vol.~19, 2006.

\bibitem{li2025reembedding}
Z.~Li, Y.~Chen, T.~Zhang, and X.~Wang, ``Reembedding and reweighting are needed for tail item sequential recommendation,'' in \emph{Proceedings of the ACM Web Conference}, 2025, pp. 4925--4936.

\bibitem{zhang2021model}
Y.~Zhang, D.~Z. Cheng, T.~Yao, X.~Yi, L.~Hong, and E.~H. Chi, ``A model of two tales: Dual transfer learning framework for improved long-tail item recommendation,'' in \emph{Proceedings of the ACM Web Conference}, 2021, pp. 2220--2231.

\bibitem{liu2023co}
Y.~Liu, X.~Zhang, M.~Zou, and Z.~Feng, ``Co-occurrence embedding enhancement for long-tail problem in multi-interest recommendation,'' in \emph{Proceedings of the 17th ACM Conference on Recommender Systems}, 2023, pp. 820--825.

\bibitem{kim2023melt}
K.~Kim, D.~Hyun, S.~Yun, and C.~Park, ``Melt: Mutual enhancement of long-tailed user and item for sequential recommendation,'' in \emph{Proceedings of the 46th International ACM SIGIR Conference on Research and Development in Information Retrieval}, 2023, pp. 68--77.

\bibitem{jang2020cities}
S.~Jang, H.~Lee, H.~Cho, and S.~Chung, ``Cities: Contextual inference of tail-item embeddings for sequential recommendation,'' in \emph{Proceedings of the IEEE International Conference on Data Mining}.\hskip 1em plus 0.5em minus 0.4em\relax IEEE, 2020, pp. 202--211.

\bibitem{qin2024metaga}
B.~Qin, Z.~Huang, Z.~Wu, C.~Wang, and Y.~Chen, ``Metaga: Metalearning with graph-attention for improved long-tail item recommendation,'' \emph{IEEE Transactions on Computational Social Systems}, 2024.

\bibitem{yang2023loam}
H.~Yang, Y.~Choi, G.~Kim, and J.-H. Lee, ``Loam: Improving long-tail session-based recommendation via niche walk augmentation and tail session mixup,'' in \emph{Proceedings of the 46th International ACM SIGIR Conference on Research and Development in Information Retrieval}, 2023, pp. 527--536.

\bibitem{wang2023multifdf}
B.~Wang, S.~Song, S.~Liu, and X.~Deng, ``Multifdf: Multi-community clustering for fairness-aware recommendation,'' \emph{IEEE Transactions on Computational Social Systems}, vol.~10, no.~6, pp. 2959--2970, 2023.

\bibitem{hu2022memory}
Y.~Hu, Y.~Liu, C.~Miao, and Y.~Miao, ``Memory bank augmented long-tail sequential recommendation,'' in \emph{Proceedings of the 31st ACM International Conference on Information and Knowledge Management}, 2022, pp. 791--801.

\bibitem{liu2024large}
Q.~Liu, X.~Wu, X.~Zhao, Y.~Wang, Z.~Zhang, F.~Tian, and Y.~Zheng, ``Large language models enhanced sequential recommendation for long-tail user and item,'' \emph{arXiv preprint arXiv:2405.20646}, 2024.

\bibitem{bian2023multi}
S.~Bian, X.~Pan, W.~X. Zhao, J.~Wang, C.~Wang, and J.-R. Wen, ``Multi-modal mixture of experts representation learning for sequential recommendation,'' in \emph{Proceedings of the 32nd ACM International Conference on Information and Knowledge Management}, 2023, pp. 110--119.

\bibitem{zhang2024relation}
Z.~Zhang, A.~Wang, Y.~Zhang, Y.~Ren, W.~Li, B.~Wang, and M.~Inuiguchi, ``Relation pruning and discriminative sampling over knowledge graph for long-tail recommendation,'' \emph{Information Sciences}, p. 120871, 2024.

\bibitem{bai2022constitutional}
Y.~Bai, S.~Kadavath, S.~Kundu, A.~Askell, J.~Kernion, A.~Jones, A.~Chen, A.~Goldie, A.~Mirhoseini, C.~McKinnon \emph{et~al.}, ``Constitutional ai: Harmlessness from ai feedback,'' \emph{arXiv preprint arXiv:2212.08073}, 2022.

\bibitem{guo2024controllable}
Y.~Guo, G.~Cui, L.~Yuan, N.~Ding, Z.~Sun, B.~Sun, H.~Chen, R.~Xie, J.~Zhou, Y.~Lin \emph{et~al.}, ``Controllable preference optimization: toward controllable multi-objective alignment,'' in \emph{Proceedings of the 2024 Conference on Empirical Methods in Natural Language Processing}, 2024, pp. 1437--1454.

\bibitem{wang2024arithmetic}
H.~Wang, Y.~Lin, W.~Xiong, R.~Yang, S.~Diao, S.~Qiu, H.~Zhao, and T.~Zhang, ``Arithmetic control of llms for diverse user preferences: directional preference alignment with multi-objective rewards,'' in \emph{Proceedings of the 62nd Annual Meeting of the Association for Computational Linguistics}.\hskip 1em plus 0.5em minus 0.4em\relax Association for Computational Linguistics (ACL), 2024, pp. 8642--8655.

\bibitem{jaques2017sequence}
N.~Jaques, S.~Gu, D.~Bahdanau, J.~M. Hern{\'a}ndez-Lobato, R.~E. Turner, and D.~Eck, ``Sequence tutor: Conservative fine-tuning of sequence generation models with kl-control,'' in \emph{Proceedings of the International Conference on Machine Learning}.\hskip 1em plus 0.5em minus 0.4em\relax PMLR, 2017, pp. 1645--1654.

\bibitem{rafailov2024direct}
R.~Rafailov, A.~Sharma, E.~Mitchell, C.~D. Manning, S.~Ermon, and C.~Finn, ``Direct preference optimization: Your language model is secretly a reward model,'' \emph{Advances in Neural Information Processing Systems}, vol.~36, 2024.

\bibitem{peters2010relative}
J.~Peters, K.~Mulling, and Y.~Altun, ``Relative entropy policy search,'' in \emph{Proceedings of the AAAI Conference on Artificial Intelligence}, vol.~24, no.~1, 2010, pp. 1607--1612.

\bibitem{peng2019advantage}
X.~B. Peng, A.~Kumar, G.~Zhang, and S.~Levine, ``Advantage-weighted regression: Simple and scalable off-policy reinforcement learning,'' \emph{arXiv preprint arXiv:1910.00177}, 2019.

\bibitem{maddison2016concrete}
C.~J. Maddison, A.~Mnih, and Y.~W. Teh, ``The concrete distribution: A continuous relaxation of discrete random variables,'' \emph{arXiv preprint arXiv:1611.00712}, 2016.

\bibitem{kool2019stochastic}
W.~Kool, H.~Van~Hoof, and M.~Welling, ``Stochastic beams and where to find them: The gumbel-top-k trick for sampling sequences without replacement,'' in \emph{Proceedings of the International Conference on Machine Learning}.\hskip 1em plus 0.5em minus 0.4em\relax PMLR, 2019, pp. 3499--3508.

\bibitem{krichene2022sampled}
W.~Krichene and S.~Rendle, ``On sampled metrics for item recommendation,'' \emph{Communications of the ACM}, vol.~65, no.~7, pp. 75--83, 2022.

\bibitem{radford2021learning}
A.~Radford, J.~W. Kim, C.~Hallacy, A.~Ramesh, G.~Goh, S.~Agarwal, G.~Sastry, A.~Askell, P.~Mishkin, J.~Clark \emph{et~al.}, ``Learning transferable visual models from natural language supervision,'' in \emph{Proceedings of the International Conference on Machine Learning}.\hskip 1em plus 0.5em minus 0.4em\relax PMLR, 2021, pp. 8748--8763.

\bibitem{hu2024enhancing}
J.~Hu, W.~Xia, X.~Zhang, C.~Fu, W.~Wu, Z.~Huan, A.~Li, Z.~Tang, and J.~Zhou, ``Enhancing sequential recommendation via llm-based semantic embedding learning,'' in \emph{Companion Proceedings of the ACM Web Conference}, 2024, pp. 103--111.

\end{thebibliography}

\vfill

\end{document}